\documentclass{aa}
\usepackage{txfonts}
\usepackage{graphicx}
\usepackage{natbib}
\bibpunct{(}{)}{;}{a}{}{,}

\begin{document}

\newcommand{\be}{\begin{equation}}
\newcommand{\ee}{\end{equation}}
\newcommand{\bdm}{\begin{displaymath}}
\newcommand{\edm}{\end{displaymath}}
\newcommand{\bea}{\begin{eqnarray}}
\newcommand{\eea}{\end{eqnarray}}

\title{Proper motions of radiative knots in simulations of stellar jets}
\subtitle{An alternative to pulsating inflow conditions}

\author{
        F. Rubini       \inst{1}
\and    S. Lorusso      \inst{1} 
  \thanks{\emph{Now at:} GE Oil \& Gas, Nuovo Pignone, Firenze, Italy}
\and    L. Del Zanna    \inst{1}
\and    F. Bacciotti    \inst{2}
}

\institute{
Dipartimento di Astronomia e Scienza dello Spazio,
Universit\`a di Firenze, Largo E. Fermi 2, 50125 Firenze, Italy
\\ \email{rubini@arcetri.astro.it}
\and
INAF - Osservatorio Astrofisico di Arcetri, 
Largo E. Fermi 5, 50125 Firenze, Italy
}

\date{Received ...; accepted ...}

\authorrunning{F. Rubini et al.}
\titlerunning{Proper motions of radiative knots in simulations of stellar jets}

\abstract
{}
{Elongated jets from young stellar objects typically present a
nodular structure, formed by a chain of bright \emph{knots}
of enhanced emission with individual proper motions. Though it is 
generally accepted that internal shocks play an important role in the 
formation and dynamics of such structures, their precise origin and 
the mechanisms behind the observed proper motions is still a matter of debate. 
Our goal is to study numerically the origin, dynamics, and emission 
properties of such knots.}
{Axisymmetric simulations are performed with a shock-capturing code for
gas dynamics, allowing for molecular, atomic, and ionized hydrogen in
non-equilibrium concentrations subject to ionization/recombination
processes. Radiative losses in [\ion{S}{II}] lines are computed, and
the resulting synthetic emission maps are compared with observations.}
{We show that a pattern of regularly spaced internal oblique shocks,
characterized by individual proper motions, is generated by the pressure 
gradient between the propagating jet and the time variable external cocoon.
In the case of under-expanded, light jets the resulting emission knots
are found to move downstream with the jet flow, with increasing velocity and
decaying brightness toward the leading bow shock. 
This suggests that the basic properties of the knots observed in stellar 
jets can be reproduced even without invoking \emph{ad hoc} pulsating 
conditions at the jet inlet, though an interplay between the two scenarios 
is certainly possible.
}
{}
   

\maketitle
%

\section{Introduction}

Collimated outflows and jets seem to be ubiquitous features in
astrophysics and are observed over a wide range of spatial scales,
from several megaparsecs for extragalactic sources (AGNs) down to
a few parsecs for young stellar objects (YSOs). In both cases a chain
of bright \emph{knots} is typically observed along the jet, suggesting
that similar physical processes may be at work in spite of the enormous
difference in scale and, most probably, even in the jet composition itself.
As far as YSO jets are concerned, the bright knots observed in emission
lines (also known as Herbig-Haro, HH, objects) represent spectacular 
tracers for these structures over a wide range of wavelengths, and
their ubiquity suggests that they are a key component of the jet dynamics. 

Though widely observed in detailed images and spectra 
\citep[e.g.][]{reipurth01}, the precise nature and origin of the knots is,
however, still the subject of investigation. In the first bright 
section of an optical jet, up to $\approx 0.1$~pc from the source, 
the number of knots ranges typically between 5 and 20. Their pattern 
is not static, but moves at a substantial fraction, from $70\%$ or more, 
of the flow speed, a value increasing with distance from the source
\citep{eisloeffel92}. On the other hand, the knots brightness tends to 
decay along the jet axis \citep{morse92,ray96,reipurth01}.
Several jets have been imaged at high resolution ($\la 0.1\arcsec $) 
using the Hubble Space Telescope (HST) \citep{ray96,reipurth97,hartigan01,
bally02,reipurth02,hartigan05}. As an example, HST images of HH 46/47 exhibit 
a complex structure in which the [\ion{S}{II}] emission decouples 
from the \ion{H}{$\alpha$} emission \citep{heathcote96}. 
Both lines define a chain of small knots with spacing of $2-3\arcsec$.
The \ion{H}{$\alpha$} - [\ion{S}{II}] difference image shows that the 
[\ion{S}{II}] emission is more extended along the flow than in
\ion{H}{$\alpha$} emission maps, where filaments and wisps are visible,
produced either at the front of internal working surfaces or in shock
induced local changes in the flow direction. 
In this framework the \ion{H}{$\alpha$} emission is produced in the hotter
and more excited thin layer just behind the shock front, while 
[\ion{S}{II}] lines come from a denser and cooler layer more distant 
from the front, which has a larger extension \citep{bacciotti99a}.  
HH~47 also shows a set of leading bow shocks with spacing much larger than 
the intervals between neighboring knots. 

The current interpretation for the origin of both the beam \emph{mini} 
bow shocks and the large leading bow shocks involves the presence 
of pulsation in the ejection mechanism 
\citep{raga92,stone93,falle95,suttner97,raga98,raga02} 
(the \emph{working surfaces} scenario). The physical reason for such 
pulsation, however, has yet not been clearly identified. 
It is possible that some kind of self-consistent MHD mechanism is
at work \citep[e.g.][]{ouyed97}.
A pulsating inflow has also been applied to 3-D simulations 
which include the effects of precession due to the rotation
of the nozzle \citep{cerqueira04,cerqueira06}, motivated by the recent
observations of toroidal velocities at the base of jets associated
with T~Tauri stars \citep{bacciotti02b,coffey04,woitas05,coffey07}. 
Finally, the effects of magnetic fields in axisymmetric simulations 
with periodically varying inflow conditions have been also studied 
\citep{cerqueira97,osullivan00,stone00,massaglia05,decolle06}.

On the other hand, other HH jets such as HH~30 \citep{ray96,bacciotti99b} 
do not present bow shock-like features in the forbidden lines, 
and the knots look more like axially symmetric \emph{blobs}
aligned with the jet axis. It has been suggested in the past that 
this appearance can be generated by Kelvin-Helmholtz instabilities. 
In particular, \citet{bodo94} have investigated the growth of 
Kelvin-Helmoltz instabilities in a Cartesian slab of supersonic, 
adiabatic flow interacting with the external matter at rest. 
In their paper the authors show that as these instabilities 
grow linearly and eventually saturate, a diamond-like pattern
of internal oblique shocks (IOS) is formed. However, the periodic boundary 
conditions imposed on the longitudinal jet propagation direction
prevent all effects due to the non-stationary jet propagation and 
necessarily yield IOS patterns which are static with respect to the
mean flow.

In the present study we perform numerical simulations 
of cooling jets which originate from a nozzle, to recover the effects 
of non-stationary jet propagation, but we \emph{do not} impose a 
periodically variable inflow velocity as it is generally assumed. 
Encouraged by our preliminary results \citep{rubini04}, where intermittent  
patterns of knots were found in some particular cases even without 
\emph{ad hoc} assumptions on the inflow speed, our intention here is to 
study, in greater detail and for a wider choice of parameters, the
formation mechanism, kinematics, and emission properties of the IOS
which are seen to form in the region behind the jet head. These arise 
due to the pressure gradient between the jet and the external medium, 
and may be ultimately responsible for the observed knotty emission.
3-D effects, such as those due to precession motions at the nozzle 
or to kink-like instabilities induced by magnetic field, are 
expected to be more important far away from the source. For this reason,
a simple axisymmetric hydrodynamical model can be used when investigating
the region near the nozzle, where knots are seen to form and appear
still well collimated with the jet beam. Moreover,
a cooling model based on a three-species network (neutral, ionized,
and molecular hydrogen) can be safely assumed since we are not interested
in resolving the small cooling scales behind the leading bow-shock,
where $\sim 10-20$ species are typically evolved. For recent results
on this subject see \citet{raga07} and references therein.

The steady inflow scenario had been investigated
by other authors, but it was argued that such IOS should be smoothed 
away in radiative jets by the cooling losses \citep[e.g.][]{blondin90}. 
\citet{cerqueira97} also remarked that in magnetized jets the cooling
tends to smooth out pinch-like instabilities. In general, it is 
agreed that radiative losses tend to reduce the shock strength 
\citep{downes98,micono98,micono00}, so the idea has been put aside 
in favor of a pulsating inflow.
However, by using the most recent indications concerning the physical 
parameters of the gas at the jet base, obtained from spectral diagnostics
of high angular resolution data \citep[e.g.][]{lavalley00,bacciotti02a,
hartigan04}, it is possible to determine the initial conditions of the 
numerical simulations in a more realistic way than has been done 
in the past.
Here we are able to show that IOS are able to survive the cooling losses and to
form a regular pattern of emitting knots on length scales which
mostly depend on the pressure ratio between jet and interstellar 
medium. Moreover, such a pattern is not static, but the individual knots
show a degree of proper motion even when steady inflow conditions are imposed.
This result is basically due to the fact that the gas surrounding the 
jet beam, the \emph{cocoon}, is a highly dynamic and 
time-varying environment. Thus, the interpretation of the optical knots
as being due to IOS cannot be ruled out and most probably cooperates
with the other effects successfully proposed so far.

The paper is structured as follows. In Sect.~2 we discuss the equations,
the numerical method, and the general mechanism for the formation of IOS
and for their proper motion. In Sect.~3 we present various results 
from the simulations, including a comparison with observations.
Section~4 is devoted to the final discussions.
                                                           
\section{The physical and numerical model}

\subsection{Gas dynamic equations and source terms}
In our simulations we solve the system of hydrodynamical fluid equations
for a cooling gas formed by molecular ($\mathrm{H}_2$), atomic ($\ion{H}{I}$), 
and ionized ($\ion{H}{II}$) hydrogen in non-equilibrium concentrations,
plus atomic helium ($\mathrm{He}$) and heavier elements in fixed concentrations
and free electrons ($\mathrm{e}$).
The evolution equations for such a system, in conservation form as 
required by shock-capturing numerical schemes, are the following:
\be
\frac{\partial \rho }{\partial t}+\vec{\nabla} \cdot (\rho \vec{u})=0,
\label{eq:cont}
\ee
\be
\frac{\partial }{\partial t}(\rho\vec{u})+\vec{\nabla} \cdot 
(\rho \vec{u}\vec{u}+p\vec{I})=\vec{0},  \label{eq:moto}
\ee
\be
\frac{\partial }{\partial t}(\rho E)+\vec{\nabla} \cdot (\rho H\vec{u}
)=-\mathcal{S}_\mathrm{rad}-\mathcal{S}_\mathrm{ion}-
\mathcal{S}_{\mathrm{H}_2},  \label{eq:ener}
\ee
\be
\frac{\partial N_\ion{H}{II}}{\partial t}+\vec{\nabla} \cdot 
(N_\ion{H}{II}\,\vec{u})=\mathcal{N}_\mathrm{ion}-\mathcal{N}_\mathrm{rec},   
\label{eq:nhii}
\ee
\be
\frac{\partial N_{\mathrm{H}_2}}{\partial t}+\vec{\nabla} \cdot 
(N_{\mathrm{H}_2}\,\vec{u})=
-\mathcal{N}_\mathrm{diss}.   \label{eq:nh2}
\ee
In the above equations $\rho$, $\vec{u}$ and $p$ represent density, velocity
and pressure for each volume element, $\vec{I}$ is the identity tensor,
$E=\varepsilon +\frac{1}{2}u^{2}$ is internal plus kinetic energy per
unit mass, and $H=E+p/\rho $ is the specific enthalpy. 
These evolution equations are completed by the relation for the total
number density
\be
N_\mathrm{tot}=N_\ion{H}{II}+N_\ion{H}{I}+N_{\mathrm{H}_2}
+N_\mathrm{e}+N_\mathrm{He}, \label{eq:ntot}
\ee
which is used to derive the gas temperature, where $N_\ion{H}{II}$, 
$N_\ion{H}{I}$, $N_{\mathrm{H}_2}$, $N_\mathrm{e}$, and $N_\mathrm{He}$ are
the number densities for ionized, atomic (neutral), molecular hydrogen, 
electrons, and helium atoms, respectively. The source terms on the right 
hand side of Eq.~(\ref{eq:ener}), the energy equation, are respectively: 
$\mathcal{S}_\mathrm{rad}$, which contains all radiative losses excepting
those due to $\mathrm{H}_2$; $\mathcal{S}_{\mathrm{H}_2}$, where radiative 
and dissociation losses from $\mathrm{H}_2$ are included; 
$\mathcal{S}_\mathrm{ion}$, which takes into account 
hydrogen ionization. These contributions, which require computation
of the particle densities, will be discussed later. Note that
the non-equilibrium hypothesis leads to the two extra continuity 
equations, namely Eq.~(\ref{eq:nhii}) for protons and Eq.~(\ref{eq:nh2}) 
for molecular hydrogen. In the former, the term 
$\mathcal{N}_\mathrm{ion}-\mathcal{N}_\mathrm{rec}$ is the proton
population variation, per unit of time and volume, due to ionization and
recombination, whereas in the latter $\mathcal{N}_\mathrm{diss}$ 
is the molecular hydrogen density decrease rate due to dissociation processes. 
Reformation processes have not been considered here, 
since densities for molecular hydrogen reformation are too small, and 
reformation time scales are too large compared with 
dissociation time scales.
Solar abundances of heavier elements in fixed concentrations have been 
also included, since they are important for energy losses though
are too small to affect the mass.

For the computation of the total particle density in Eq.~(\ref{eq:ntot}),
which is needed to update the temperature field, the following steps 
are required:
\begin{itemize}
\item the total mass density $\rho$ and the number densities $N_\mathrm{e}$ 
and $N_{\mathrm{H}_2}$ are evolved directly via the respective continuity 
equations, thus are known quantities at each time-step;
\item let us first introduce the total number density of hydrogen atoms as
\be
N_\mathrm{H}=N_\ion{H}{II}+N_\ion{H}{I}+2N_{\mathrm{H}_2}, \label{eq:nh}
\ee
including all possible states.
From the relations $\rho_\mathrm{H}=N_\mathrm{H}m_\mathrm{p}$,
$\rho_\mathrm{He}=4N_\mathrm{He}m_\mathrm{p}$ ($\rho=\rho_\mathrm{H}+
\rho_\mathrm{He}$
and $m_\mathrm{p}$ is the proton mass), and assuming helium to be present 
in a fixed concentration with $\rho_\mathrm{He}/\rho=1/4$, we easily find
\be
N_\mathrm{H}=\textstyle{\frac{3}{4}}(\rho /m_\mathrm{p}),~~~
N_\mathrm{He}=\textstyle{\frac{1}{16}}(\rho/m_\mathrm{p});
\ee
\item the number density of electrons $N_\mathrm{e}$ is derived from
\be
N_\mathrm{e}=N_\ion{H}{II}+N_\mathrm{em}=N_\ion{H}{II}+10^{-4}N_\mathrm{H},
\ee
where $N_\mathrm{em}$ is the concentration of free electrons due to metals, 
again assumed to be a fixed fraction of $N_\mathrm{H}$;
\item the number density of atomic hydrogen $N_\ion{H}{I}$ is 
finally derived from Eq.~(\ref{eq:nh}).
\end{itemize}

Once we have the temperature field, we are ready to compute all source
terms in the fluid equations. Let us discuss them in detail.
The source term in Eq.~(\ref{eq:nhii}) for ionized hydrogen is provided by 
recombination and ionization rates, respectively given by
\be
\mathcal{N}_\mathrm{rec}=N_\mathrm{e}N_\ion{H}{II}\,\alpha (T),~~~
\mathcal{N}_\mathrm{ion}=N_\mathrm{e}N_\ion{H}{I}\,\beta (T).
\ee
The functions $\alpha$ and $\beta$ of temperature 
$T$ (in Kelvin degrees) are
\be
\alpha (T)=2.06\times 10^{-11} T^{-1/2}\phi~~\mbox{cm}^3~~\mbox{s}^{-1}, 
\label{eq:alpha}
\ee
\be
\beta (T)=7.80\times 10^{-11} T^{1/2}\exp (-13.6\,
\mbox{eV}/k_{B}T)~~\mbox{cm}^3 \,\mbox{s}^{-1}, \label{eq:beta}
\ee
where $\phi$ in Eq.~(\ref{eq:alpha}) is a decreasing function of the 
temperature, ranging approximately from 4 to 1 in the region of interest 
(see Table 5.2 in \citet{spitzer78}), whereas Eq.~(\ref{eq:beta}) is taken 
from \citet{lang75} for hydrogen neutral atoms mainly in the ground level, 
as expected for a low density gas, and $k_{B}$ is the Boltzmann constant.
This yields an energy loss  contribution $\mathcal{S}_\mathrm{ion}=
(\mathcal{N}_\mathrm{ion}-\mathcal{N}_\mathrm{rec})\cdot 13.6~\mbox{eV}$ 
due to ionization of hydrogen in Eq.~(\ref{eq:ener}).

The source term in the molecular hydrogen equation arises from molecular 
dissociation alone, due to collisions with H$_2$, \ion{H}{I}, \ion{H}{II} 
and electrons. This yields
\be
\mathcal{N} _\mathrm{diss}=N_{\mathrm{H}_{2}}[
N_{\mathrm{H}_{2}}K_{\mathrm{H}_{2}}+N_\ion{H}{I}K_\ion{H}{I}
+N_{\ion{H}{II}}K_\ion{H}{II}+N_\mathrm{e}K_\mathrm{e}],
\ee
in which dissociation rates $K_{\mathrm{H}_{2}}$ and $K_\ion{H}{I}$ 
due to collisions with neutrals, H$_{2}$ and \ion{H}{I}, respectively,
are computed through the model by \citet{lepp83},
whereas those due to collisions with charged particles, $K_\ion{H}{II}$
and $K_\mathrm{e}$, derive from \citet{hollenbach89} and \citet{maclow86}. 
Corrections to the critical density in order to separate high and low 
density regimes, as suggested by \citet{martin96}, have also been introduced.
The corresponding contribution to the total energy source is
$\mathcal{S}_{\mathrm{H}_{2}}=\Lambda_\mathrm{rad}+\Lambda_\mathrm{diss}$, 
where $\Lambda_\mathrm{diss}=\mathcal{N}_\mathrm{diss}\cdot 4.48~\mbox{eV}$ 
comes from dissociations while $\Lambda _{rad}$ is mainly due to 
collisions with \ion{H}{I} and H$_{2}$ \citep[see][]{lepp83}.

Finally, we consider the cooling function $\mathcal{S}_\mathrm{rad}$.
For $T>10^{4}$~K we use the cooling function as in
\citet{dalgarno72}, which considers radiative losses due to de-excitation 
of \ion{H}{I} energy levels. Here we assume the gas to be optically thin.
For $T<10^{4}$~K the cooling function is extended by computing energy
losses mainly due to collisions between electrons and \ion{C}{II}, \ion{O}{I},
\ion{N}{I}, \ion{Fe}{II}, \ion{O}{II}, \ion{S}{II}, or, for still lower 
temperatures, between \ion{H}{I} and \ion{C}{II}, \ion{O}{I}, \ion{Si}{II},
\ion{Fe}{II} \citep{bacciotti95}. Moreover, we must consider that 
in star-forming regions, refractory species concentrate at the surface 
of dust grains. The cooling function has been corrected accordingly
to take into account reduced concentrations with respect 
to standard metal abundances \citep{sofia94}.

\subsection{Simulation setup and choice of parameters}

%
%
\begin{table}
\centering
\begin{tabular} {ccccccccc}
\hline
Case & $ \eta$ & $\Pi$ & $r_\mathrm{jet}/\bar{L}$ &
 $x_\mathrm{e}^\mathrm{jet}$ & $x_{\mathrm{H}_2}^\mathrm{jet}$ &
 $x_\mathrm{e}^\mathrm{ISM}$ & $x_{\mathrm{H}_2}^\mathrm{ISM}$  \\
\hline
$\mathcal{A}$ & 10   & 600 &0.4 & 0.3 & 0. & 0.& 0.40 \\ 
$\mathcal{B}$ &  1   &  60 &0.4 & 0.3 & 0. & 0.& 0.40 \\ 
$\mathcal{C}$ &  0.1 &  6  &0.4 & 0.3 & 0. & 0.& 0.40 \\ 
$\mathcal{D}$ &  0.4 & 24  &0.1 & 0.3 & 0. & 0.& 0.25 \\   
$\mathcal{E}$ &  0.4 & 24  &0.1 & 0.3 & 0. & 0.& 0.25 \\ 
\hline
\end{tabular}
\caption{Run parameters. According to our definitions,  $\rho_\mathrm{jet}=\eta\, 0.836\times 10^{-20}\mbox{ g cm}^{-3}$, $p_\mathrm{jet}=\Pi\, 1.381\times 10^{-10}\mbox{ erg cm}^{-3}$.  The jet velocity is always assumed $V_\mathrm{jet}= 200\mbox{ km s}^{-1}$. Case $\mathcal{E}$ is the same as $\mathcal{D}$ but with restarting inflow conditions.}
\label{tab:1}
\end{table}

The numerical code used in our simulations is a finite volume
Godunov-type scheme which solves Eqs.~(\ref{eq:cont})-(\ref{eq:nh2}),
in 2-D, by assuming axisymmetry around the jet axis and adopting 
cylindrical coordinates $(\xi,r)$, where $\xi$ is the distance from
the source along the jet and $r$ is the radius. We use second order limited 
reconstruction on characteristic variables and an exact Riemann 
solver at cell interfaces to work out numerical fluxes, which in our 
simulations has turned out to be more robust with respect to a Roe solver.
The jet originates from a \emph{nozzle} of a given radius, located at the 
left side of the numerical box and then propagates in the $\xi$ direction.
Convergence tests have been performed to find the optimal numerical 
parameters which reconcile accuracy and efficiency. 
A stretched grid with 150 points is used in the radial direction, 
which uses 20 points to describe the nozzle region, 
from $r=0$ to $r=r_\mathrm{jet}$,
whereas along $\xi$ we expand the numerical box by
adding points (up to 200 at most), while maintaining a fixed resolution,
as the jet propagates in the unperturbed medium. 
Details of the numerical method
and tests may be found in \citet{lorusso99}.

The physical parameters to be initialized at the nozzle and in the 
external, unperturbed interstellar medium (ISM), are density, pressure
(and, consequently, temperature), ionization and molecular hydrogen 
fractions (defined as $x_\mathrm{e}\simeq x_\ion{H}{II}=
N_\ion{H}{II}/(N_\ion{H}{I}+N_\ion{H}{II})$ and $x_{\mathrm{H}_2}=
N_{\mathrm{H}_2}/N_\mathrm{H}$, respectively), 
plus jet radius $r_\mathrm{jet}$ and inflow velocity $V_\mathrm{jet}$.
Most of these parameters can be derived from observations, 
either directly or combined with spectral diagnostics at moderate and
high angular resolution \citep{bacciotti99a,lavalley00,bacciotti02b}.
Typical observed tangential velocities are about $100-200\mbox{ km s}^{-1}$
and we assume $V_\mathrm{jet} = 200\mbox{ km s}^{-1}$ (which corresponds to a 
Mach number of the order of 20 for the assumed temperature, see below). 
In general, at $100-200$~AU from 
the source, the ionization fraction $x_e$ and the electron density 
$N_e$ in the jet are in the range $0.2-0.5$ and $10^2-10^3\mbox{ cm}^{-3}$, 
respectively, while the temperature is in the range $6\,000-10\,000$~K. 
From this point onward the chain of knots begins to be visible and the  
physical quantities mentioned above oscillate around the values 
attained at 200 AU from the source 
\citep[see, e.g., Figs.~1 and 2 in][]{bacciotti99b}. 
According to these analyses we place the inlet at about 200~AU from the star.

The precise parameters for all runs are shown in Table~\ref{tab:1}.
Rather than giving the jet density and pressure, we fixed only the ISM
values and we prescribed the two quantities 
$\eta = \rho_\mathrm{jet}/\rho_\mathrm{ISM}$
and $\Pi=p_\mathrm{jet}/p_\mathrm{ISM}$. 
Namely, the jet density and pressure are derived from the given 
values assuming $\rho_\mathrm{ISM}=0.5\bar{\rho}$ and 
$p_\mathrm{ISM}=\bar{p}$. Normalization is against $\bar{L}=10^{15}$~cm, 
$\bar{\rho}=1.673\times 10^{-20}\mbox{ g cm}^{-3}$ (corresponding to 
$10^4$ hydrogen atoms per cm$^3$), 
$\bar{p}=1.381\times 10^{-10}\mbox{ erg cm}^{-3}$ (corresponding to 
a temperature of $100$~K for a purely atomic hydrogen gas at density 
$\bar{\rho}$). Here we will always assume
$\Pi=60\,\eta$, to preserve the expected high temperature ratio 
between jet and ISM, with the condition $\Pi>1$.
The choice of $\Pi$ (or $\eta$) 
plays a key role, since the pressure ratio affects dramatically
the jet behavior and, in particular, the formation of IOS.
This, in turn, leads to quite different knot patterns, as we show 
in the next subsection. Another parameter affecting the structure of
IOS, in particular their dimension and spacing, is obviously 
$r_\mathrm{jet}$. 
Note that cases $\mathcal{A}$, $\mathcal{B}$, and $\mathcal{C}$  
differ only for the density ratio $\eta$ (and thus also $\Pi$), 
which is decreased from 10 to 0.1 (thus we move from \emph{heavy} 
to \emph{light} jets). Run $\mathcal{D}$ is a light jet
case where parameters have been further optimized
to provide knots with a structure similar to that observed, whereas
run $\mathcal{E}$ retains the same parameters as in run $\mathcal{D}$, but
the ejection is turned off for a while in order to reproduce some
observations where two separate chains of knots are found (case of
\emph{restarting} jet, see Sect.~\ref{sect:restart}).

\subsection{Formation mechanism of compression regions in under-expanded jets}

%
%

\begin{figure}
\resizebox{\hsize}{!}{
\includegraphics[height=5cm,width=5cm,angle=-90]{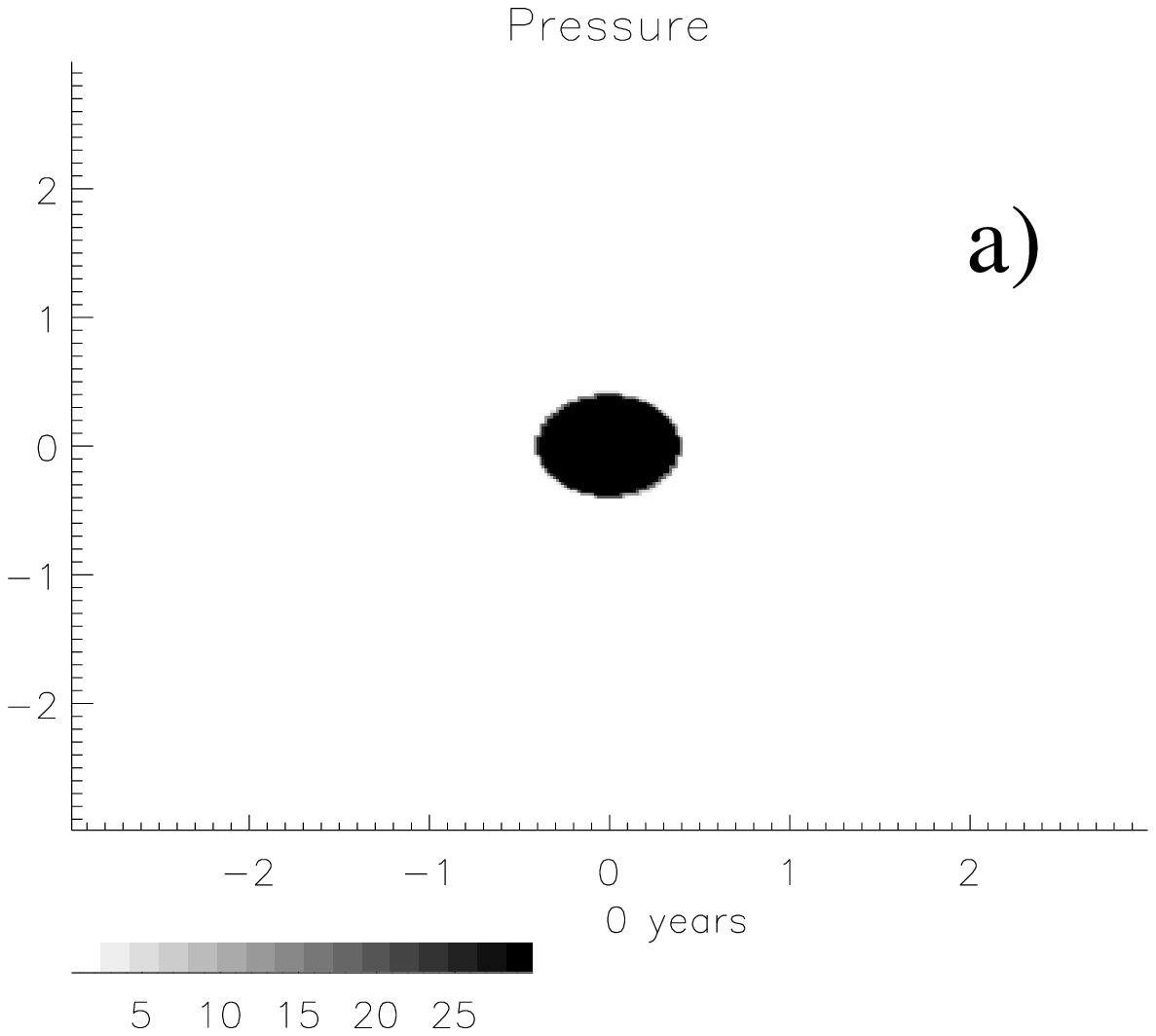}
\includegraphics[height=5cm,width=5cm,angle=-90]{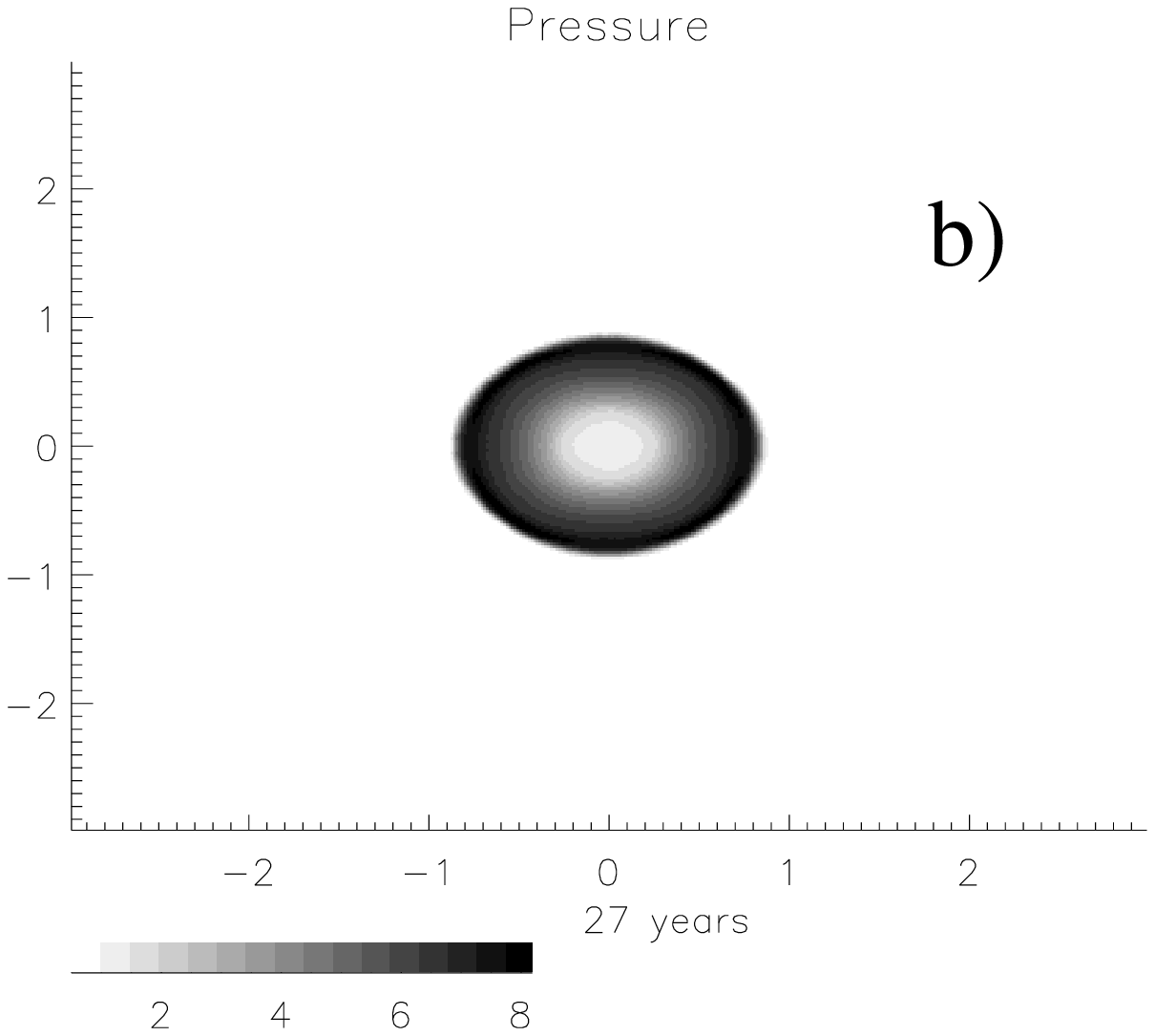}
}
\resizebox{\hsize}{!}{
\includegraphics[height=5cm,width=5cm,angle=-90]{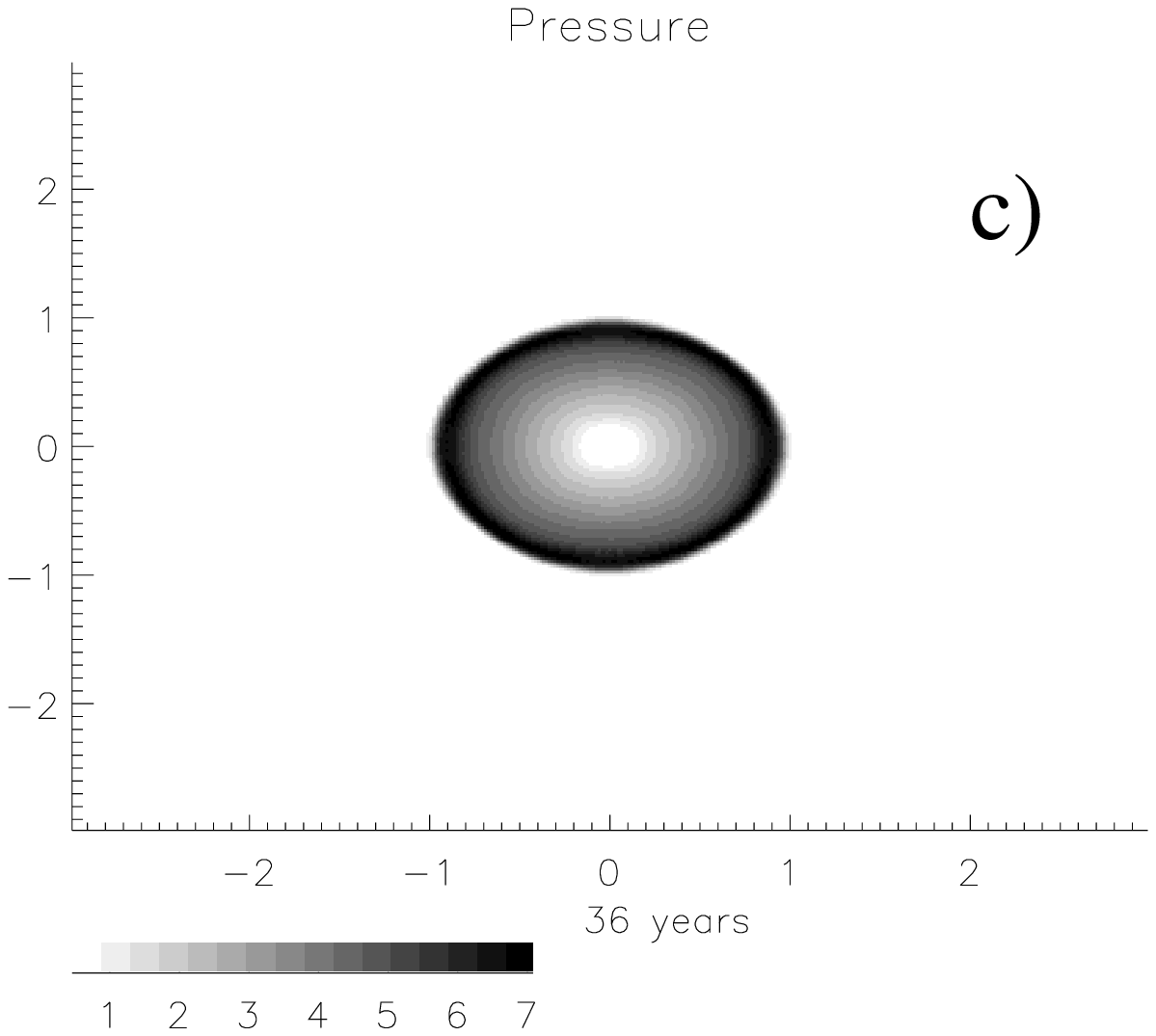}
\includegraphics[height=5cm,width=5cm,angle=-90]{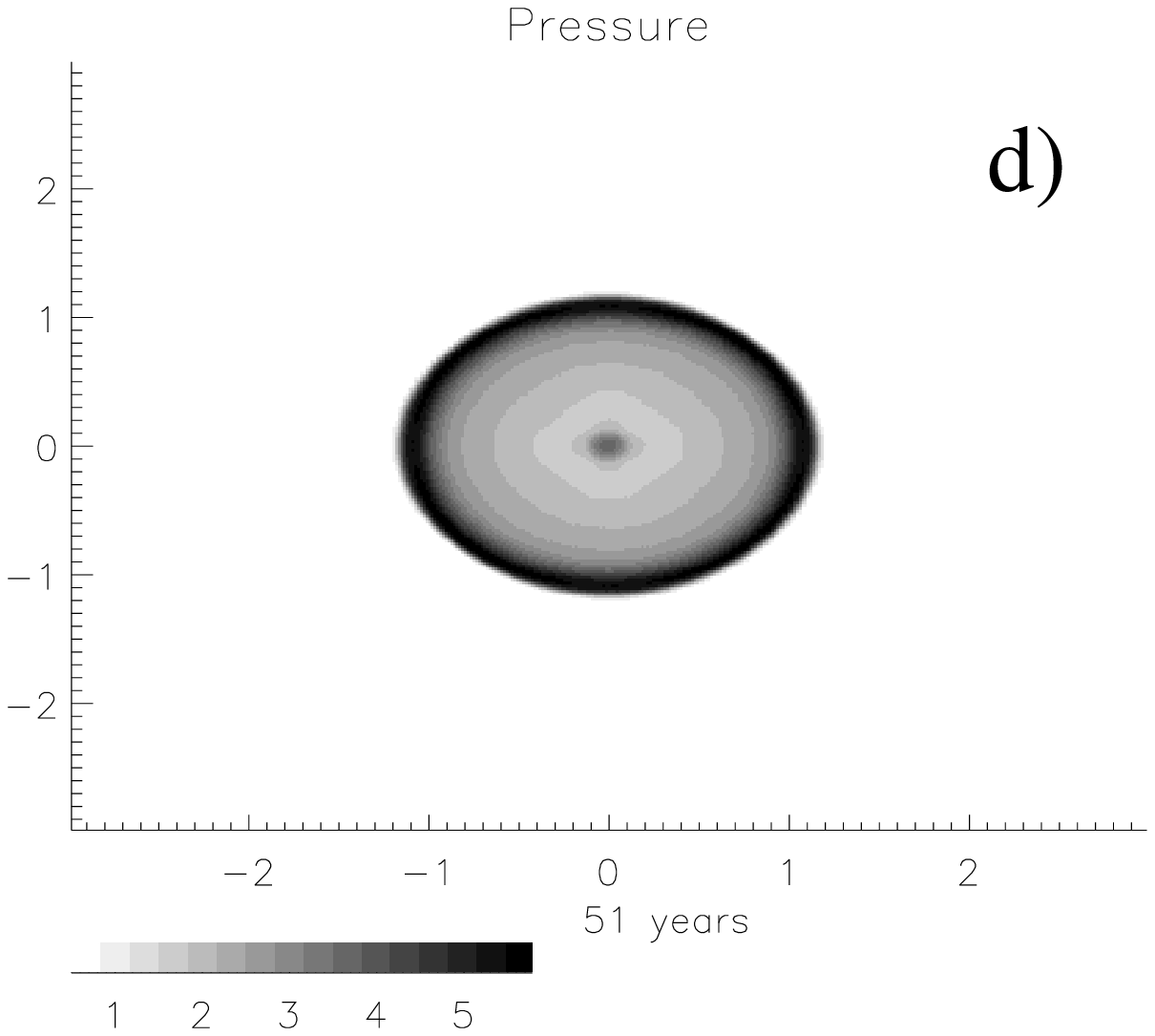}
}
\caption{Case B1. Time evolution of the pressure field.
Times are $t=0$~yr (a), $t=27$~yr (b), $t=36$~yr (c), $t=51$~yr (d). 
The pressure scale is $\bar{p}=1.381\times10^{-10}\mbox{ erg cm}^3$,
while here the length scale is $10^{14}$~cm. 
The lower (white) limit in the color bar is $0.5$, corresponding to 
the value assumed for the external ISM pressure.}
\label{fig:f1}
\end{figure}

%
%
\begin{figure}
\centering
\includegraphics[height=7cm,width=7cm,angle=-90]{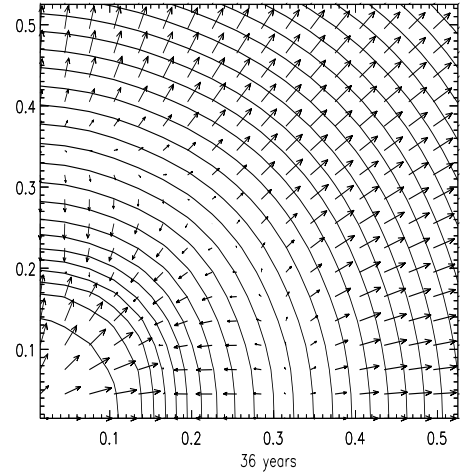}
\caption{Case B1. Velocity vectors at $t=36$~yr. A shell of 
gas is moving toward the axis pushed by the inward pressure gradient,
while the external shells are still expanding outward.}
\label{fig:f2}
\end{figure}

We now investigate the formation mechanism of compression regions
and IOS in \emph{under-expanded} jets, that is when 
$P_\mathrm{jet}>P_\mathrm{ISM}\Rightarrow\Pi>1$. Under these conditions,
just after the nozzle the gas starts to expand downward 
until the pressure falls below the surrounding local pressure in the 
\emph{cocoon} environment. Outward from this pressure equilibrium point
the flow is deviated by the inward radial pressure gradient, forming blobs 
of compressed gas aligned along the axis, which will drive the formation of 
radiatively emitting knots. 
This is the formation mechanism of IOS in jets, which is well known 
in gas dynamics. Here we study how the resulting expansion and 
compression features depend on the initial parameters.

To this aim the code has been adapted to simulate the 2-D
expansion of a circular surface at constant $\xi$, under-expanded 
with respect to the environment, i.e. to the ISM. 
The circular surface mimics an expanding slice of gas 
perpendicular to the jet axis and co-moving with the jet itself. 
At time $t=0$ it represents the jet nozzle, 
which has been initialized with a typical set of inflow parameters. 
In this qualitative model the effects of the longitudinal 
propagation of the jet, 
including re-circulating flows, cocoon temporal variability 
and Kelvin-Helmoltz instabilities arising from the 
longitudinal velocity shear, 
have been neglected. In the present section, both adiabatic (A1, B1, C1) 
and radiative (A2, B2, C2) simulations 
are performed, with density ratio $\eta = 10, 1, 0.1$ 
and pressure ratio $\Pi = 600, 60, 6$, respectively. 
The other parameters 
are the same as in the runs $\mathcal{A}$, $\mathcal{B}$, $\mathcal{C}$ 
listed in Table~\ref{tab:1}, respectively,
except for the nozzle size, whose radius is ten times smaller,
$0.4 \times 10^{14}$~cm, also leading to a smaller time evolution scale. 
In the jet simulations presented in the following sections a more 
realistic beam width has been used, to match the estimated mass loss rate.

Case B1 ($\eta=1$, $\Pi = 60$, adiabatic) is chosen to 
illustrate the general features of the gas radial evolution as a function 
of time, described in Fig.~\ref{fig:f1}.
Plot (a) represents the initial condition, with uniform pressure distribution 
inside the jet beam. At $t = 27$~yr (b) the shell is expanding: a strong 
pressure wave moves outward, and an evacuated region forms, inside the beam, 
driven by the internal rarefaction wave. 
When the inward pressure gradient becomes strong enough, particles
from the intermediate shells start to move toward the center (c, $t = 36$~yr).
This reverse flow can be seen in the velocity pattern of Fig.~\ref{fig:f2}. 
Particles moving toward the axis finally generate a central peak of pressure, 
which will result in a radiatively emitting knot (Fig.~\ref{fig:f1}d).

%
%
\begin{figure}
\centering
\includegraphics[width=9cm,height=6cm]{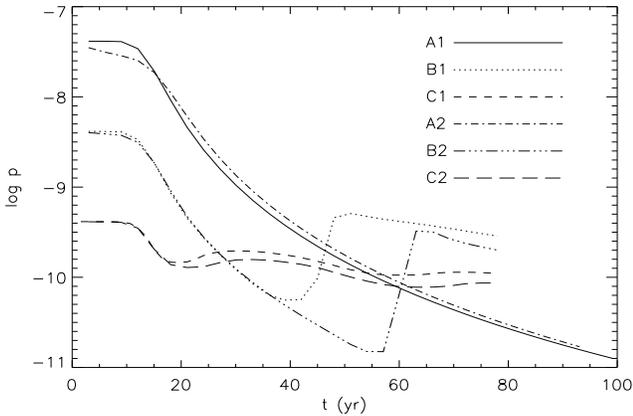}
\caption{Plots of the central pressure (in logarithmic scale) versus time 
for the three values of the pressure ratio $\Pi=600$, 60, 6 chosen for
our simulations (cases A, B, C, respectively). 
Both adiabatic (1) and radiative (2) runs are shown.
}
\label{fig:f3}
\end{figure}

The combined effects of radiative losses and different pressure ratios 
are shown in Fig.~\ref{fig:f3}, which shows the pressure taken   
on the axis versus time for all cases. Notice that the pressure jump,   
coinciding with the formation of the
central peak, does not occur at the same time for all runs. In general,
the smaller the pressure ratio $\Pi$, the smaller the knot formation 
time scale, and in turn the smaller the distance from the injection point. 
In the simulation with $\Pi = 6$, case C, a knot quite 
close to the source is produced (at $t = 30$~yr, corresponding to a
distance $L\approx 2\times 10^{16}$~cm if the particles co-moved with 
the jet at the constant velocity of $200~\mbox{km s}^{-1}$). The peak
formation is most apparent for case B, where we clearly
see a pressure jump, whereas in run A, with the highest
pressure ratio $\Pi = 600$, no knot is generated over the length scale 
corresponding to 100~yr, at the same jet velocity.
As far as radiative losses are concerned, 
these in general depend on the matter density
 and tend to delay the pressure jump and lower the peak value,
as is most apparent by comparing cases B1 and B2. 
In the radiative case A2, in particular, the jet matter 
is so overpressured and dense that radiative losses initially result 
in a quick pressure drop larger than for the adiabatic case A1. 
This radiative loss is also able to keep the pressure value at the center 
below the adiabatic value at later times.
All cases, however, show that the value of the pressure jump is not 
substantially affected by cooling losses. Instead, on-axis 
peak and minimum pressure values are smaller than for 
adiabatic simulations. Radiative losses, in fact, effectively
lower the gas adiabatic index $\gamma$, 
and allow for a denser and cooler central region. 

As already anticipated, the nozzle radius is another key parameter that
regulates the knots' length scale. 
The smaller the nozzle radius, the shorter the distance 
traveled by sound waves in the time it takes to go from the axis to 
the beam surface and back to the axis, where the compression region forms.
We have investigated the suitable initial conditions for a 
jet simulation. To summarize, under-expanded light jets with small 
jet/ambient pressure ratio (but still greater than 1), and small nozzle 
size are expected to produce nodular structures more similar to those 
observed, in terms of spacing of the knots and closeness to the source.
Moreover, contrary to common belief, the formation mechanism described 
in this section allows for \emph{proper motion} of knots in real jets. 
Unlike in this simple model, where the external pressure is constant 
with time, the beam in a stellar jet is in fact embedded in a variable 
cocoon. 
At a given point $\xi$ on the longitudinal axis the local 
pressure ratio $\Pi_\mathrm{jc}(\xi)$ between the pressure of the jet and the 
cocoon embedding the beam (which is not equal to the constant value of $\Pi$) 
actually changes in time, and the knot has to move to adjust 
the position to match the new value of $\Pi_\mathrm{jc}(\xi)$, 
according to Fig.~\ref{fig:f3}. 
Namely, $\Pi_\mathrm{jc}(\xi)$ is expected to grow with time, since     
the pressure inside the beam is steadily fed by the nozzle and stays 
unchanged, while that in the cocoon decays because of both cooling losses 
and lateral expansion. Therefore, the knot is expected to move downward 
following the gas stream, at some fraction of the injection speed, 
as is shown in next section.

\section{Simulation results and emission maps}

In the present section we show the results of the (radiative) simulations
for all the cases listed in Table~\ref{tab:1}, now restoring the full
settings for axisymmetric simulations in $(\xi,r)$. The results are 
illustrated with 2-D maps of the total density on a jet meridional section 
and, in some cases, with derived synthetic emission maps calculated for
the collisionally excited lines [\ion{S}{II}] $\lambda\lambda$ 6716, 6731.
These lines are commonly observed in YSO jets and thus the constructed maps 
represent a good test of the simulation results against the observations.
In some cases the integrated emission from a slice perpendicular to the 
jet axis (of normalized width 1~cm) will be shown as a function of the
distance from the source, and hereafter will be labeled as $E(\xi)$.
The [\ion{S}{II}] emission is calculated from the physical parameters 
determined in the simulation by using a public routine for a 5-level 
collisionally excited atom (A.~Raga, priv. comm.), assuming that all S 
atoms are ionized once and adopting a (constant) relative abundance 
$\mathrm{S}/\mathrm{H}=1.6\times 10^{-5}$ 
\citep[for details see][]{bacciotti95}.

\subsection{Heavy and density-matched jets 
(cases $\mathcal{A}$, $\mathcal{B}$)}
\label{sect:AB}

\begin{figure}[t]
\includegraphics[height=6cm,width=8.5cm]{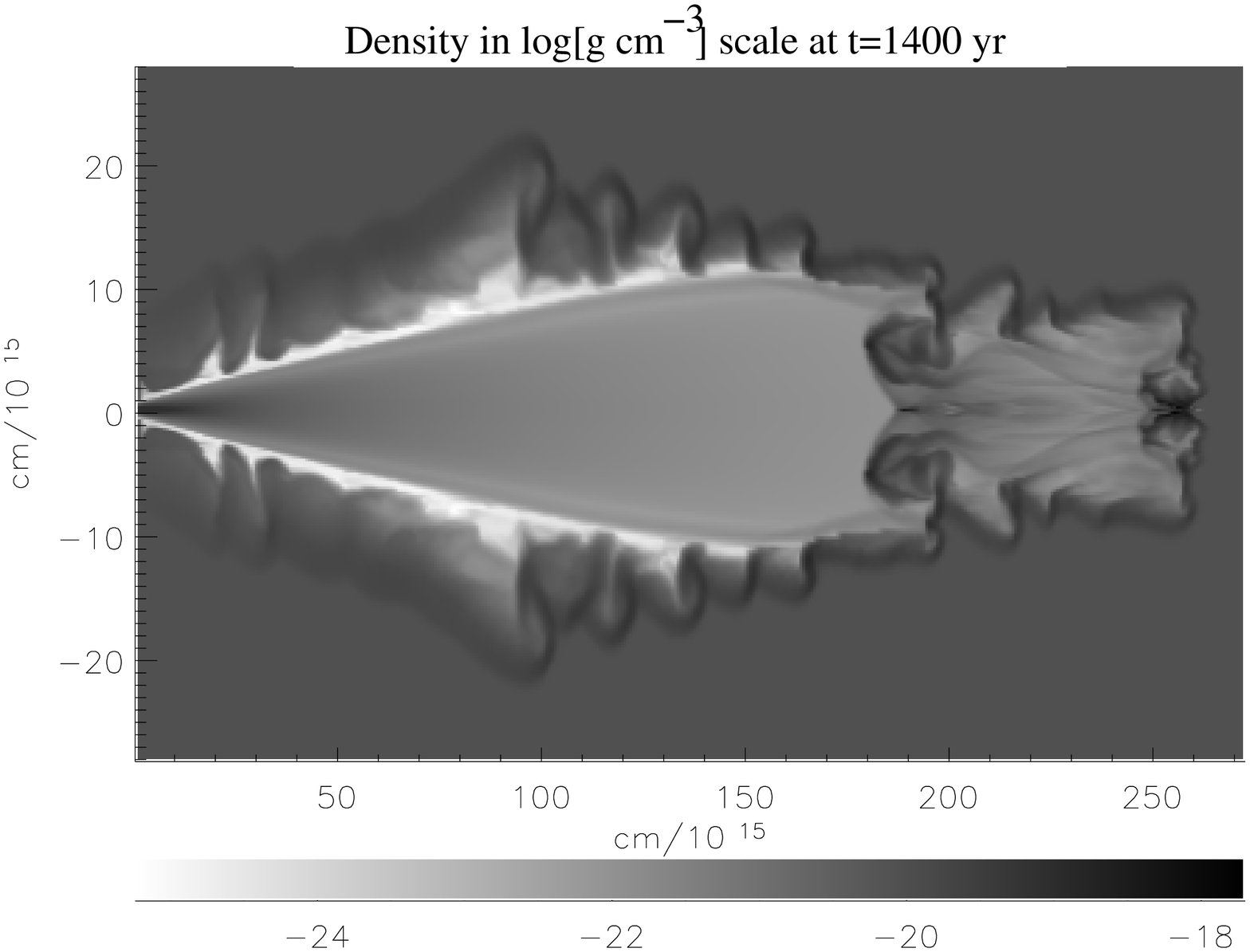}
\vskip 5mm
\hskip -2mm
\includegraphics[height=11cm]{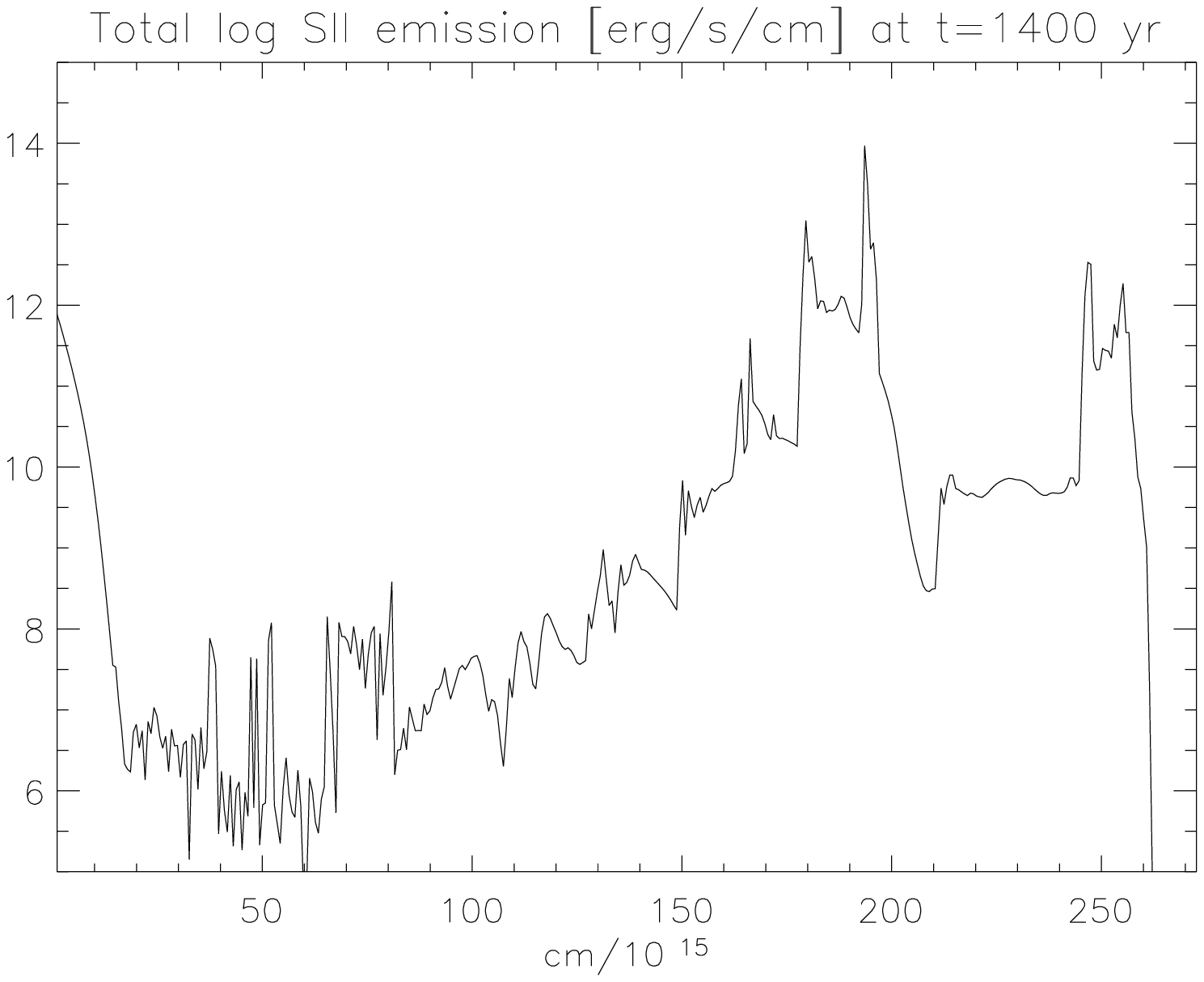}
\vskip -5.5cm
\caption{Case $\mathcal{A}$. Top: density map in logarithmic scale. 
Bottom: [\ion{S}{II}] emission $E(\xi)$, in logarithmic scale, 
integrated on a cylindrical slice (of thickness 1~cm), vs. 
the distance from the source in units of $\bar{L}$.}
\label{fig:f4}
\end{figure}

Case $\mathcal{A}$, the starting model in the parameter space of 
Table~\ref{tab:1}, represents a heavy ($\eta= 10$), strongly 
under-expanded jet with pressure ratio $\Pi = 600$.
This test case does not produce emitting knots. 
The density field (Fig.~\ref{fig:f4}, top panel) 
does not reveal any kind of internal structure. Note that in order to 
allow for a representation of some details of the jet beam, density 
and emissivity maps in the paper have an aspect ratio of the 
axes' scales far from unity. In the bottom panel
we report the integrated emission function $E(\xi)$. The figure 
only shows random discontinuities that arise from 
rings of dense matter in the external cocoon region, 
rather than from blobs of compressed gas on the axis. 
Due to the high pressure ratio at the nozzle, the gas expands 
from the origin to $\approx 150~\bar{L}$, where 
recompression occurs due to the termination shock corresponding to the 
Mach disk.
Velocity, density and pressure are rather smooth and uniform from the source 
to the Mach disk (located at $\approx 200~\bar{L}$ at $t=1400$~yr) 
and comparisons with observed images of HH objects cannot be attempted.
Downstream of the Mach disk a secondary jet forms, accelerated
by the local De Laval nozzle generated by a toroidal ring of dense
matter that forms around the Mach disk triple point \cite[e.g.][]{blondin90}.
Such an effect is frequently observed in axisymmetric HD and MHD 
simulations \citep[e.g.][]{clarke86}, while it does not appear in 
3-D simulations. 

%
%
%
%
\begin{figure}
\includegraphics[height=6cm,width=8.5cm]{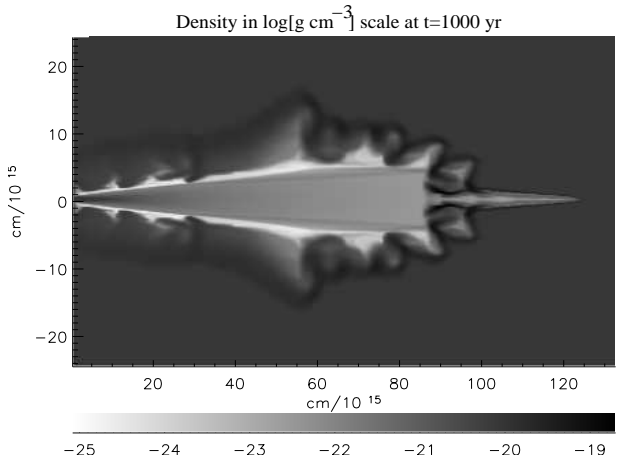}
\caption{Case $\mathcal{B}$. Density map in logarithmic scale.}
\label{fig:f5}
\end{figure}

Case $\mathcal{B}$, the density-matched run with $\eta=1$ and $\Pi = 60$,
is still far from showing satisfactory features. Density maps
are similar to those of case $\mathcal{A}$, as shown in Fig.~\ref{fig:f5}.
Although the jet is more collimated than in case $\mathcal{A}$, 
no nodular structure is visible, since the pressure ratio $\Pi$ 
at the nozzle is still too large to produce knots over the jet length. 
Note also that in this case the jet presents a narrower \emph{nose cone} 
ahead of the Mach disk.

\subsection{Light jets (case $\mathcal{C}$)}

%
%
\begin{figure*}[t]
\hskip 5mm
\includegraphics[height=7.5cm,width=6cm,angle=-90]{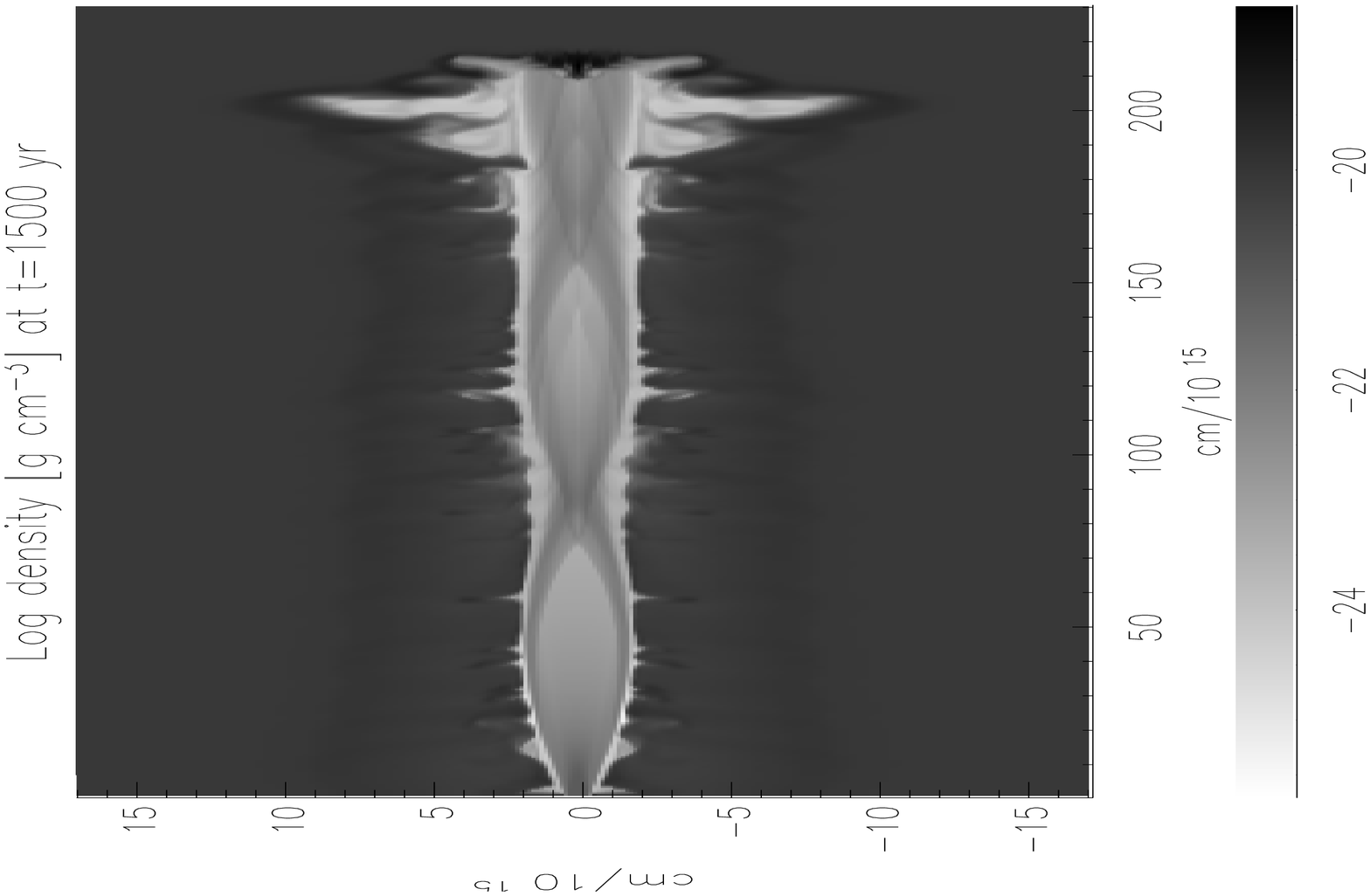}
\hskip 5mm
\includegraphics[height=8.5cm,width=6cm,angle=-90]{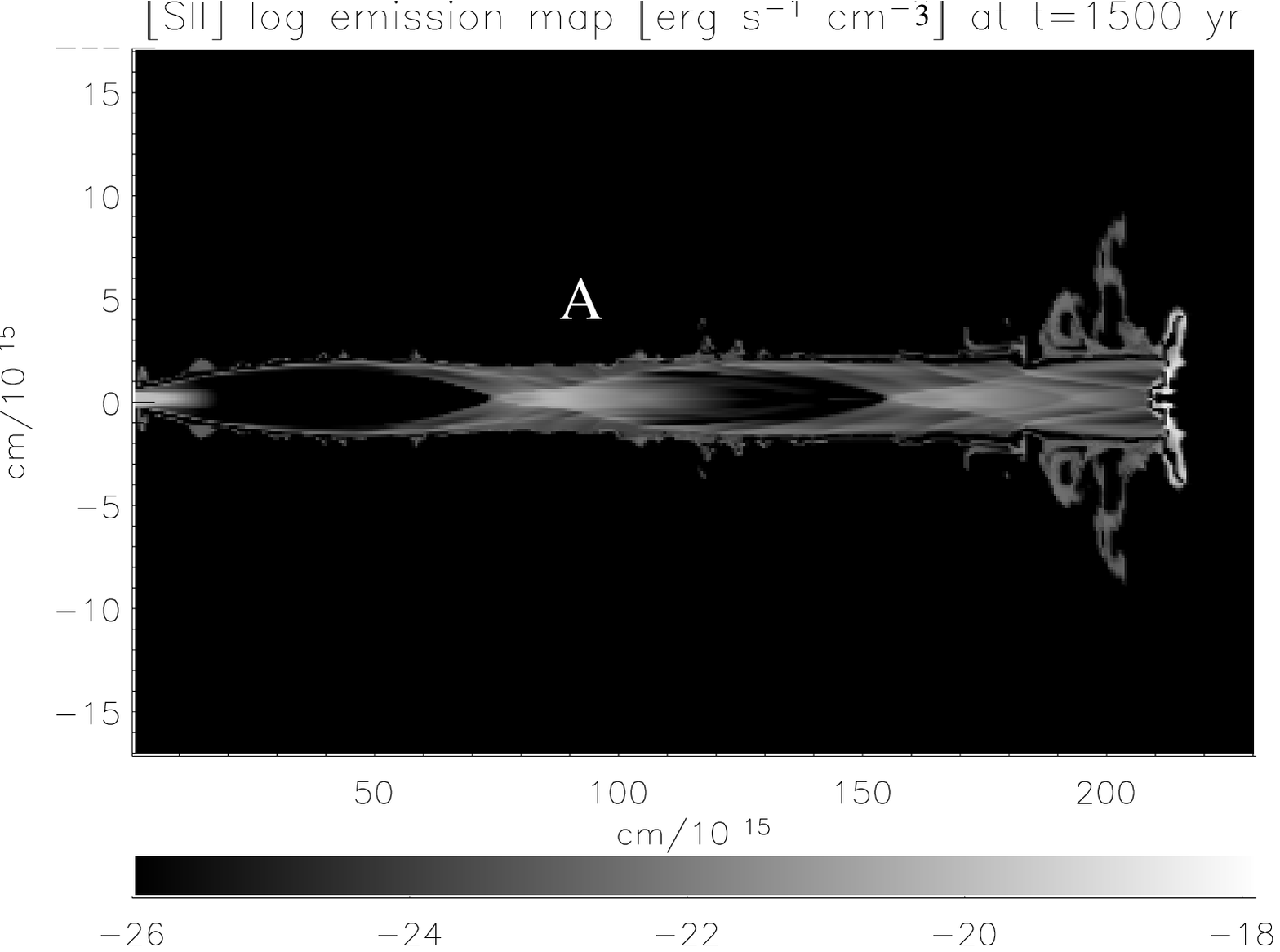}
\caption{Case $\mathcal{C}$. Left panel: density map in logarithmic scale.
Right panel: [\ion{S}{II}] emission map in logarithmic scale. 
Note the presence of a knot (A). Its estimated luminosity is 
$\sim 10^{25} \mbox{ erg s}^{-1}$.}
\label{fig:f6}
\end{figure*}

%
%
\begin{figure*}
\centering
\resizebox{\hsize}{!}{
\includegraphics[angle=-90]{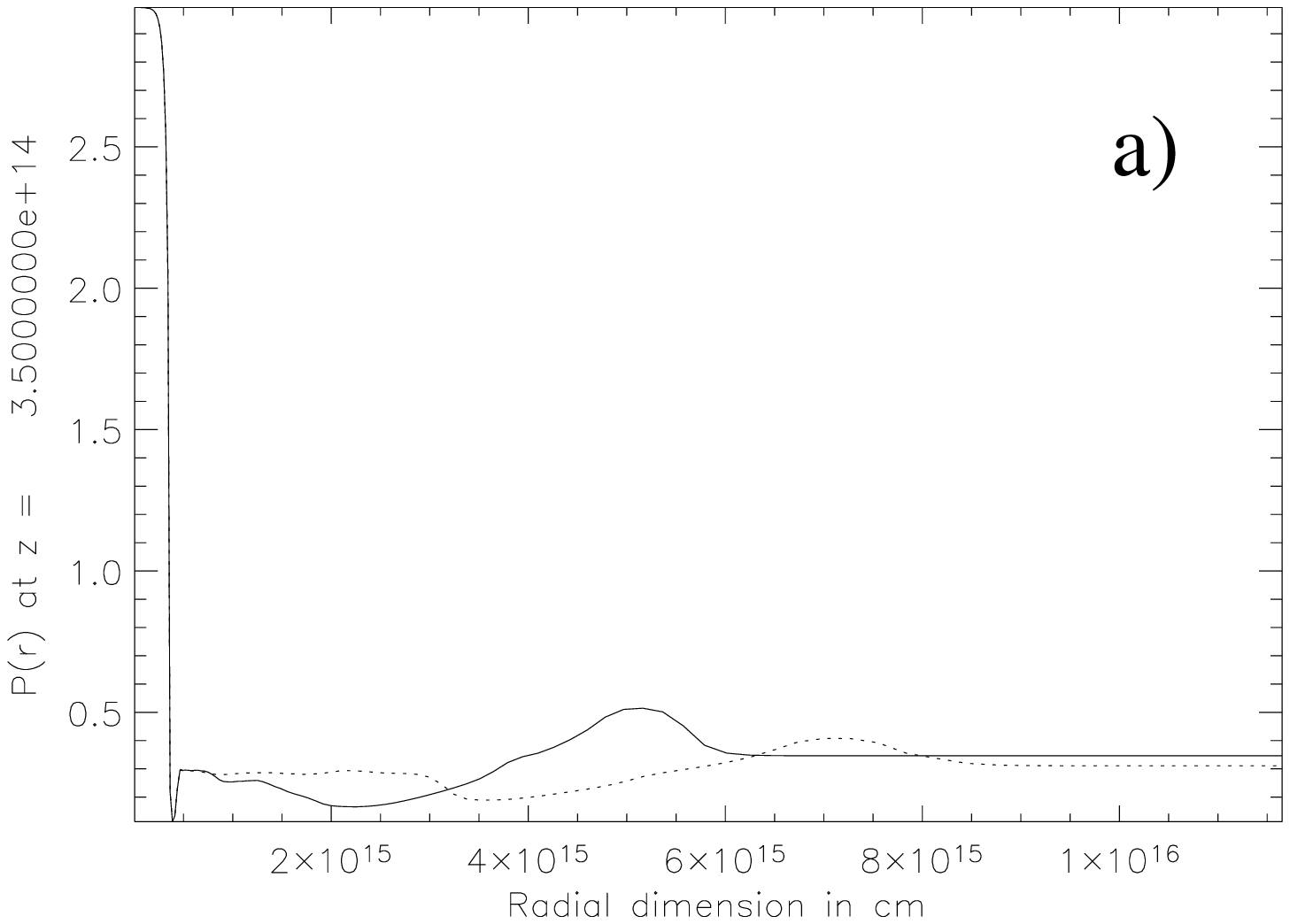}
\includegraphics[angle=-90]{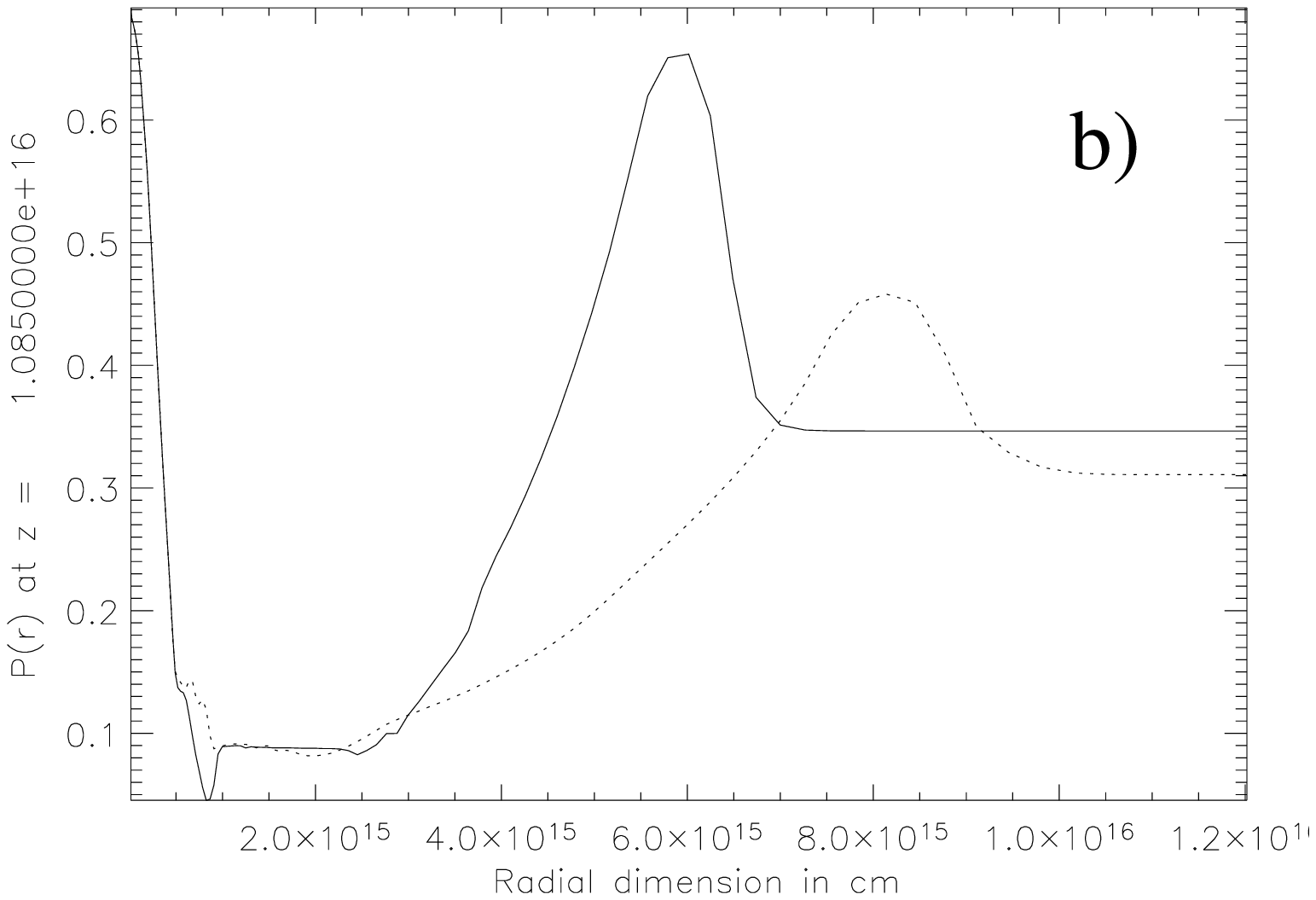}
\includegraphics[angle=-90]{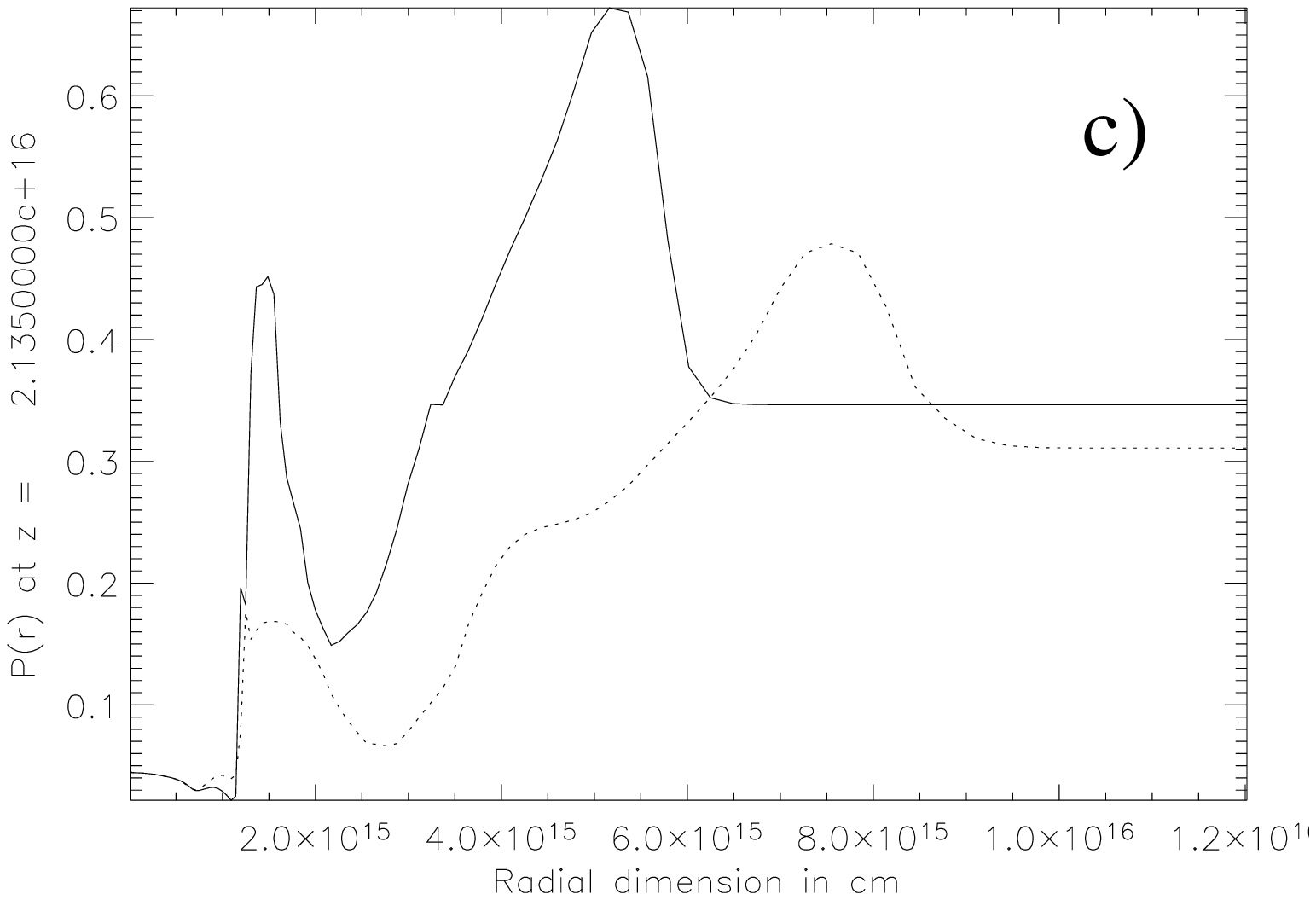}
}
\resizebox{\hsize}{!}{
\includegraphics[angle=-90]{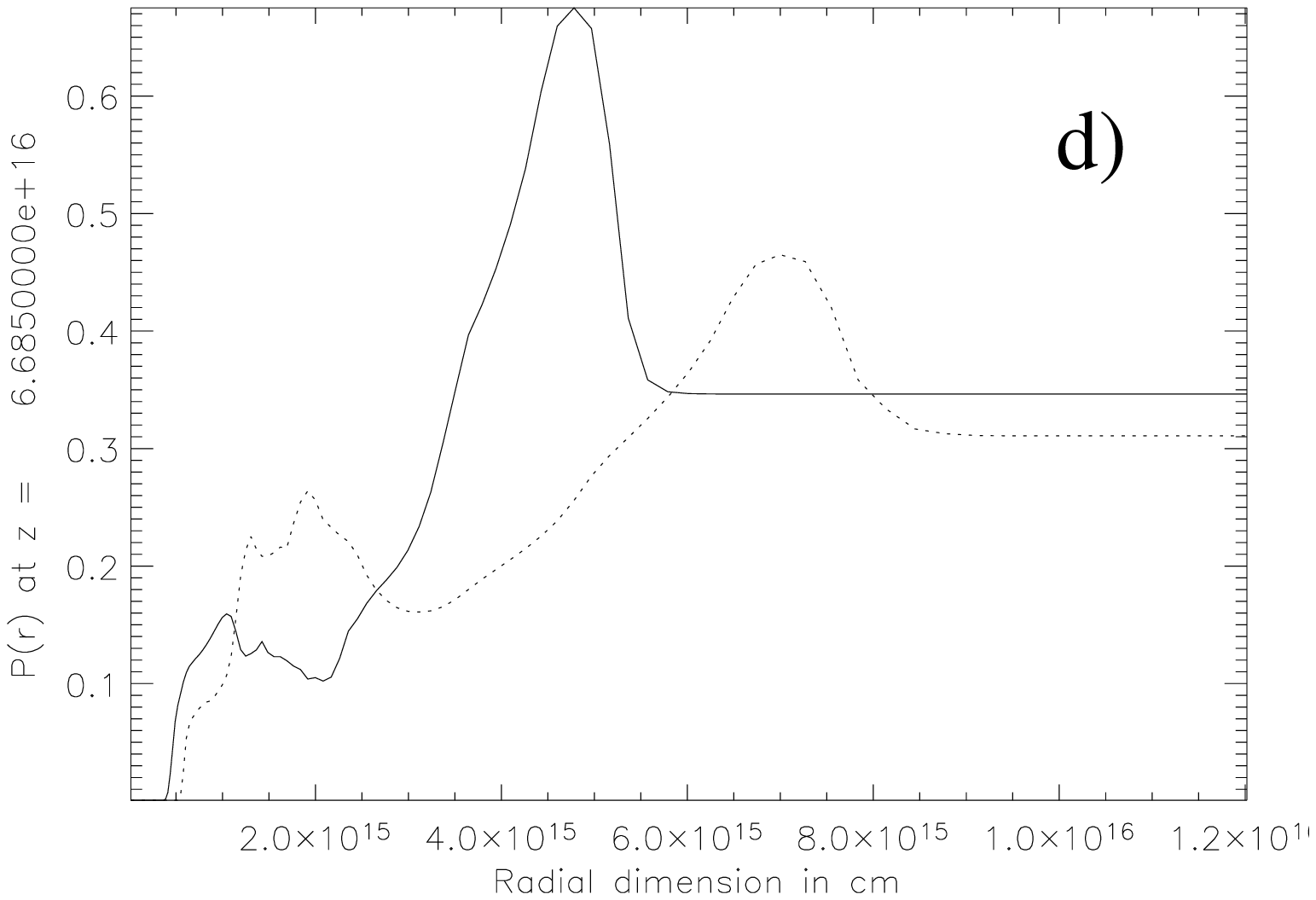}
\includegraphics[angle=-90]{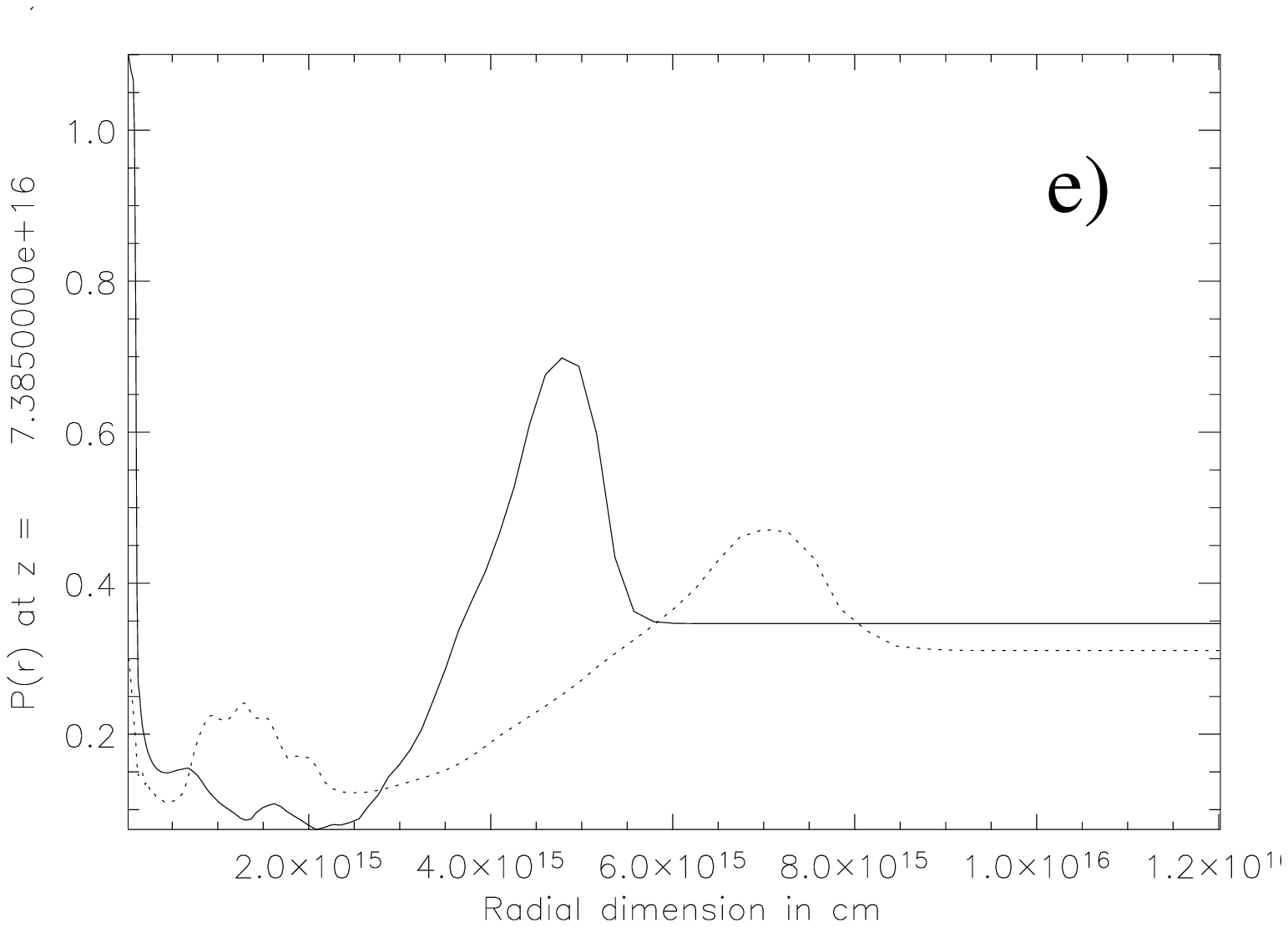}
\includegraphics[angle=-90]{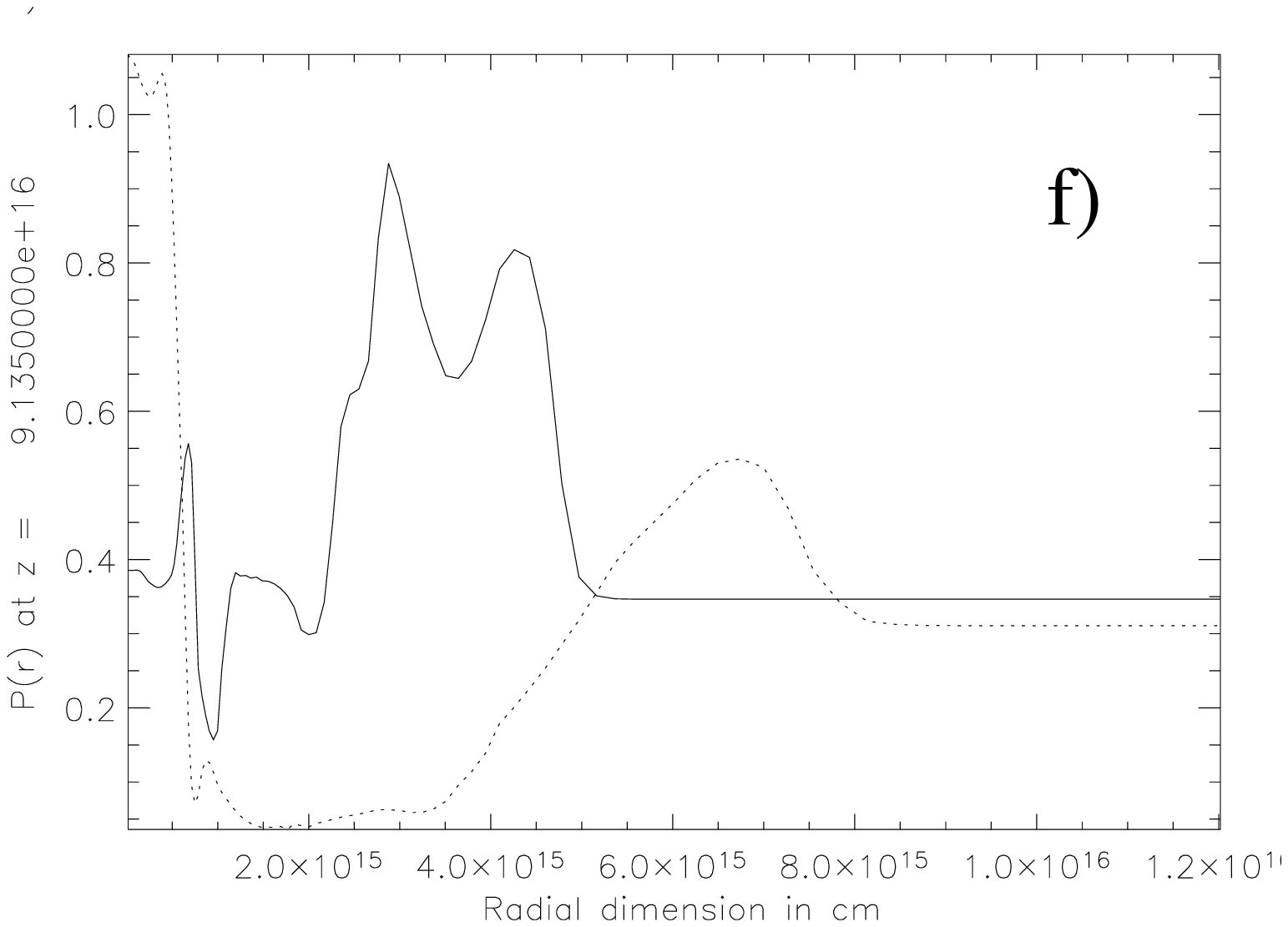}
}
\caption{Case $\mathcal{C}$. Pressure $p(r)$ as a function of
the radius, at $t = 900$~yr (solid line) and $t=1500$~yr (dotted line).
Panels (a-f) refer respectively to positions $\xi=0.35~\bar{L}$, 
$\xi=11~\bar{L}$, $\xi=21~\bar{L}$, $\xi=67~\bar{L}$, $\xi=74~\bar{L}$, 
$\xi=91~\bar{L}$. Pressure and radius scaling changes from 
picture to picture, to allow for a better representation.
}
\label{fig:f7}
\end{figure*}

In this case we have $\eta = 0.1$ and $\Pi = 6$, all other parameters 
being unchanged with respect to cases $\mathcal{A}$ and $\mathcal{B}$. 
The mass loss rate is of the order of $10^{-8}\,M_\odot\mbox{ yr}^{-1}$, 
in agreement with the estimates derived from observations 
\citep{bacciotti02b}. 
Figure~\ref{fig:f6} shows the density map (left panel) and a simulated 
image in the light of [\ion{S}{II}] lines (right panel), on the same scale
and for the same time $t=1500$~yr. A well defined knot appears at  
$\xi \approx  90~\bar{L}$ from the source (labeled with A in the
synthetic emission map), corresponding to the dark throat observed 
in the density map. The estimated luminosity of this single knot is 
$\approx 10^{25}\mbox{ erg s}^{-1}$.
This case confirms that IOS and emission knots appear for small $\Pi$ values. 
The simple model that has been used to show how blobs of compressed 
gas form in under-expanded jets, can now be checked. 
Figure~\ref{fig:f7} contains a sequence of plots showing the radial profiles
$p(r)$ of the pressure, taken at different distances $\xi$ along the jet 
axis. In each plot the solid line refers to time $t = 900$~yr, while
the dotted line refers to the same output time in Fig.~\ref{fig:f6},
$t=1500$~yr. 

\begin{figure}
\centering
\hskip -2cm
\includegraphics[width=9cm]{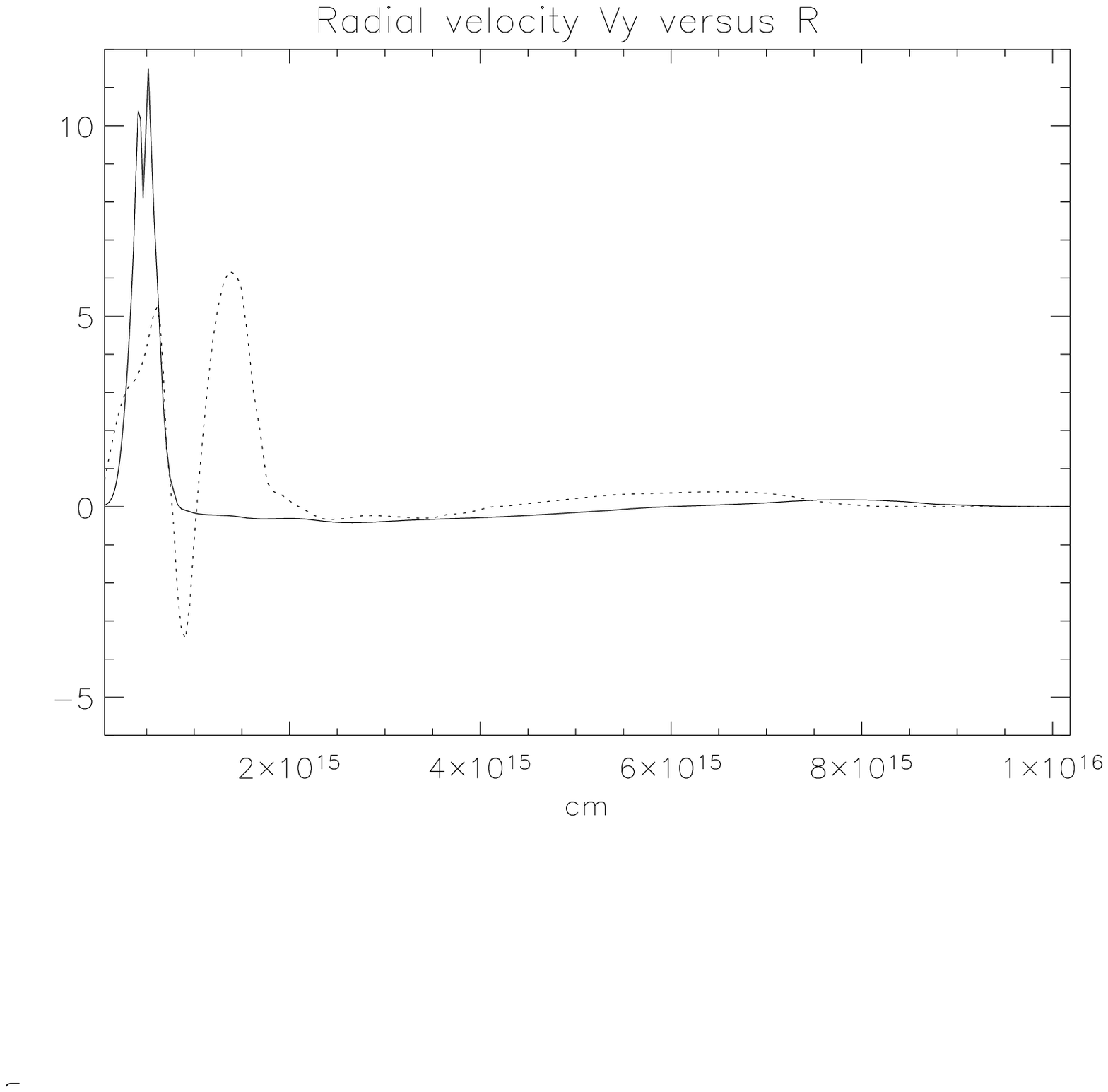}
\vskip -7cm
\caption{Case $\mathcal{C}$. Radial velocity profile $V_r(r)$ at 
$t = 900$~yr, $\xi= 11~\bar{L}$ (solid line) and $\xi=21~\bar{L}$ 
(dotted line).}
\label{fig:f8}
\end{figure}

We will now discuss the figures in some detail.
Panel (a) of Fig.~\ref{fig:f7} refers to a region quite close to the source,
$\xi=0.35~\bar{L}$. The pressure on the axis corresponds to the inflow value 
that fills the nozzle homogeneously, the external flat profile matches 
the ISM pressure. The two plots show that close to the source the pattern 
does not change in time significantly.
Panel (b) gives the situation approximately 20 nozzle radii downstream 
of the source. As particles flow down the nozzle, the beam expands 
laterally, following  the outgoing pressure wave (see Fig.~\ref{fig:f1}b 
for a comparison with the \emph{slice} model). 
Comparison between solid and dotted line in panel (b) shows that 
the beam undergoes a lateral expansion in time. At $t=1500$~yr 
the beam is wider and the average cocoon pressure is lower with respect 
to $t=900$~yr (the cocoon is defined here as the region bounded 
by the peak of outgoing pressure wave). 
Figure~\ref{fig:f7}c, to be compared to Fig.~\ref{fig:f1}c, shows the 
growth of the inward pressure wave that pushes the gas toward the axis. 
This inward flow is plotted in Fig.~\ref{fig:f8}, which shows the 
radial profile of the radial velocity $V_r(r)$, taken at the age of 
$t=900$~yr and for $\xi = 11~\bar{L}$ (quite near the source, solid line) 
and for $\xi= 21~\bar{L}$, (closer to knot A, dotted line). 
The former is overall positive (expansion phase), the latter shows that 
internal rings of matter are moving to the axis (compare also with
Fig.~\ref{fig:f2}). In panel (d) the internal compression wave is 
still moving inward, until it reaches the axis (e, see also 
Fig.~\ref{fig:f1}d). The peak of pressure on the axis in panel (e) 
corresponds to knot A in the \emph{young} jet (i.e. that at $t=900$~yr).  
The \emph{old} jet wave is still on the way and reaches the axis in panel (f), 
approximatively 40 nozzle radii downstream. This spatial delay confirms that 
as time goes by the knot forms farther and farther away from the source, 
because of the lower external pressure field, according to Fig.~\ref{fig:f3}. 
The spatial gap between the peaks in panels (e) and (f) divided by the 
time interval between what we have labeled as \emph{old} and \emph{young} 
jets measures the knot proper motion. In this case the resulting velocity 
is too small when compared to observations ($V_\mathrm{knot} \approx 9 
\mbox{ km s}^{-1}$).

%
%

\subsection{Light jets with smaller radius (case $\mathcal{D}$)} 

%
%
\begin{figure*}
\centering
\includegraphics[height=11cm,width=8cm]{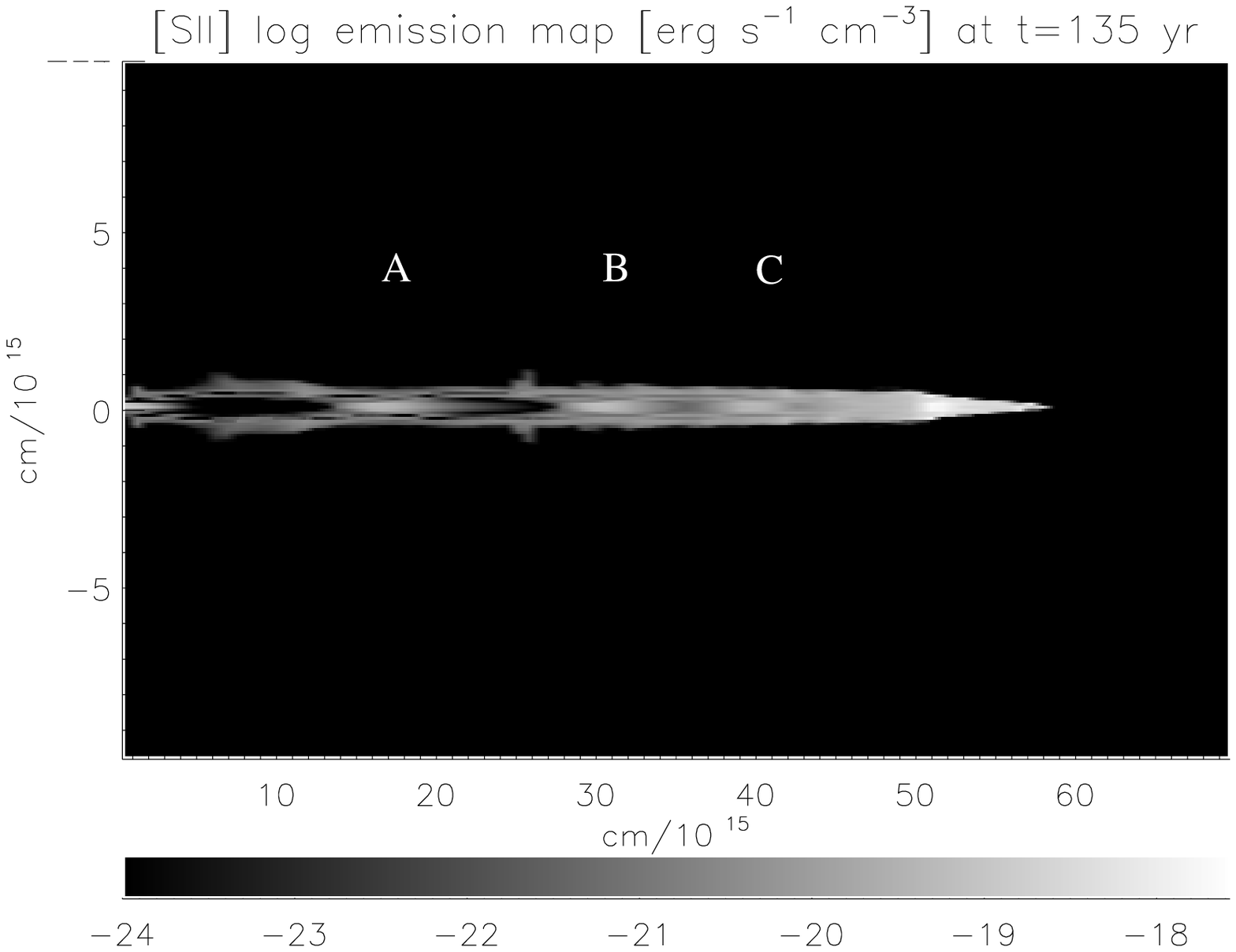}
\includegraphics[height=11cm,width=8cm]{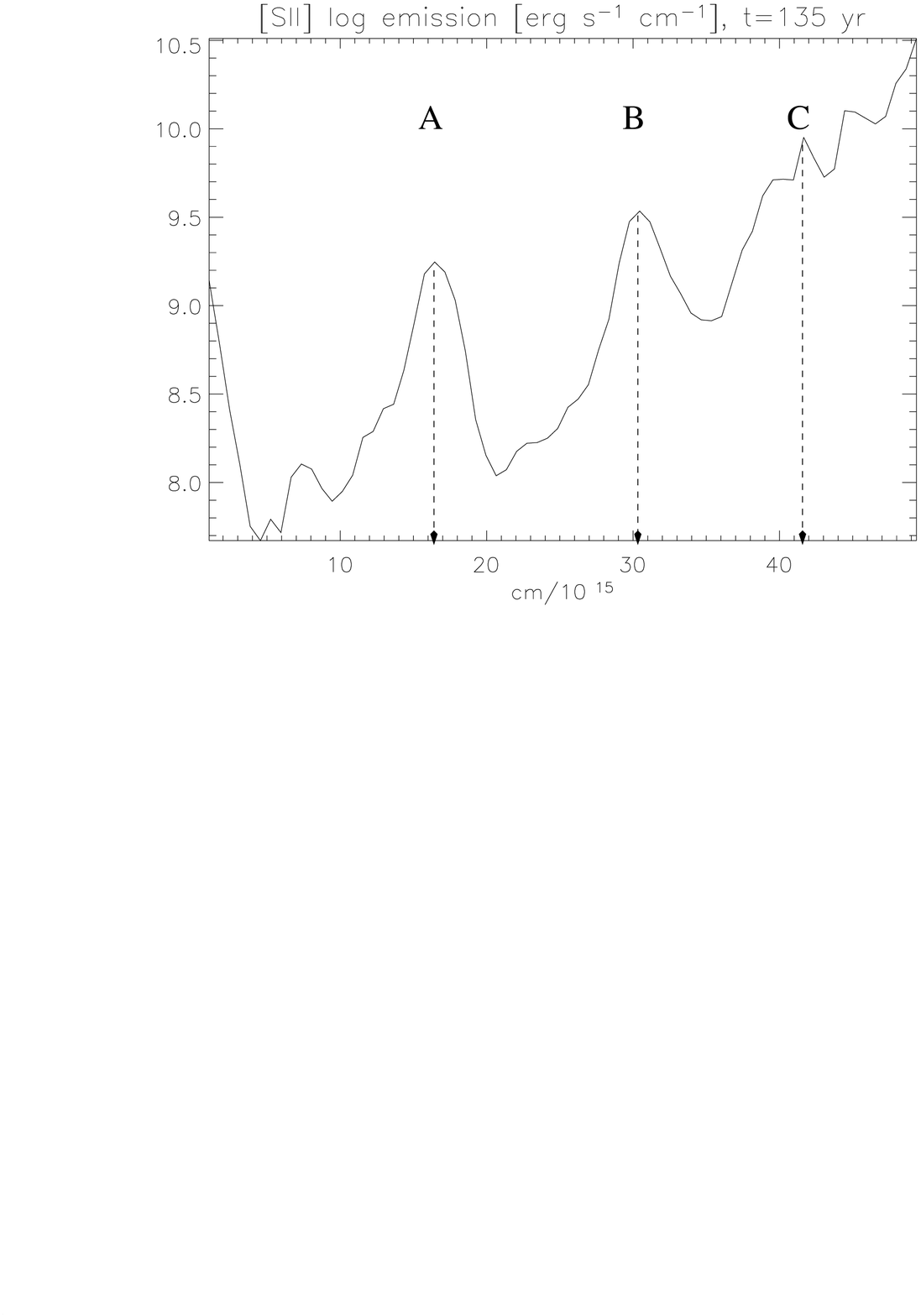} \\ 
\vskip -5cm
\includegraphics[height=11cm,width=8cm]{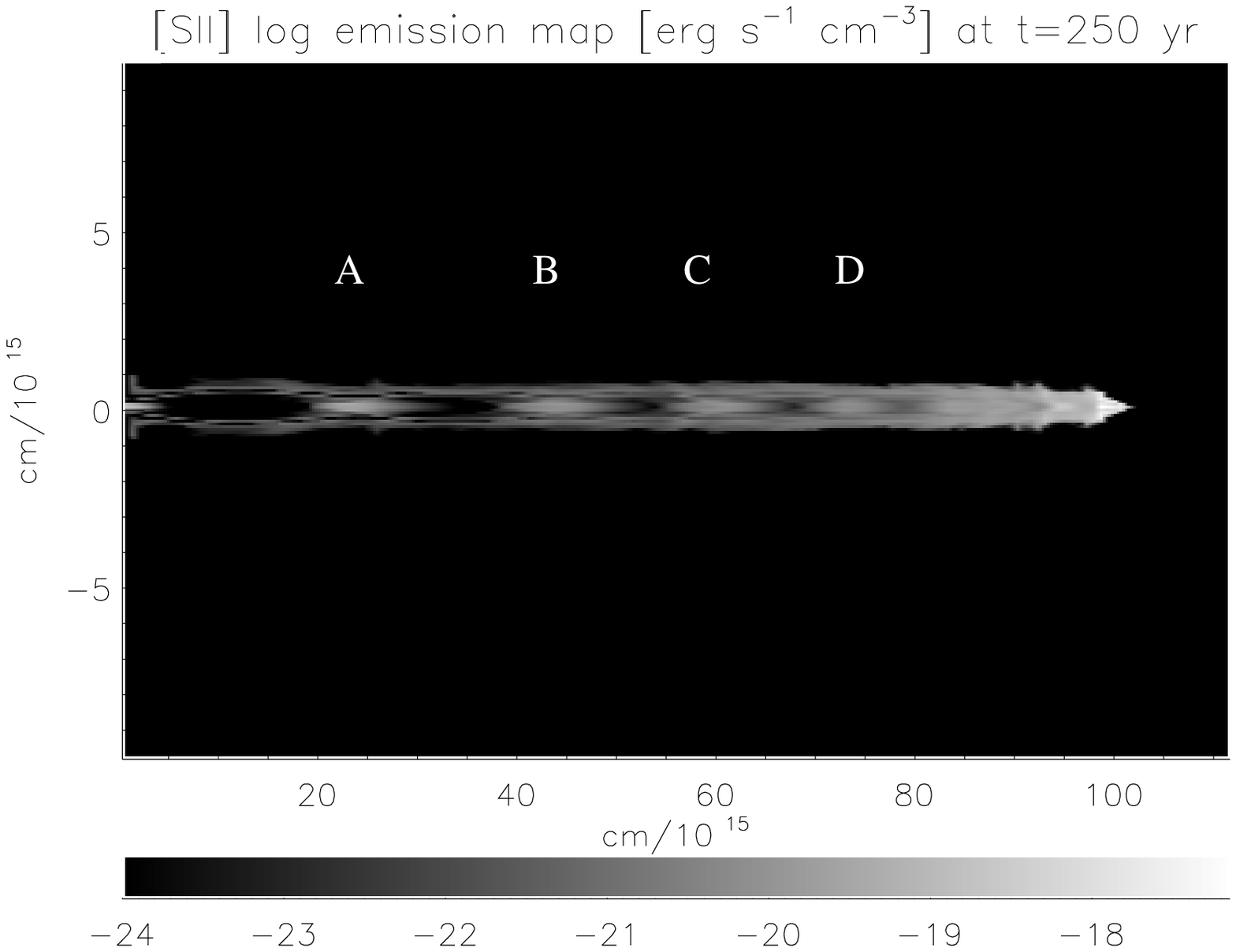}
\includegraphics[height=11cm,width=8cm]{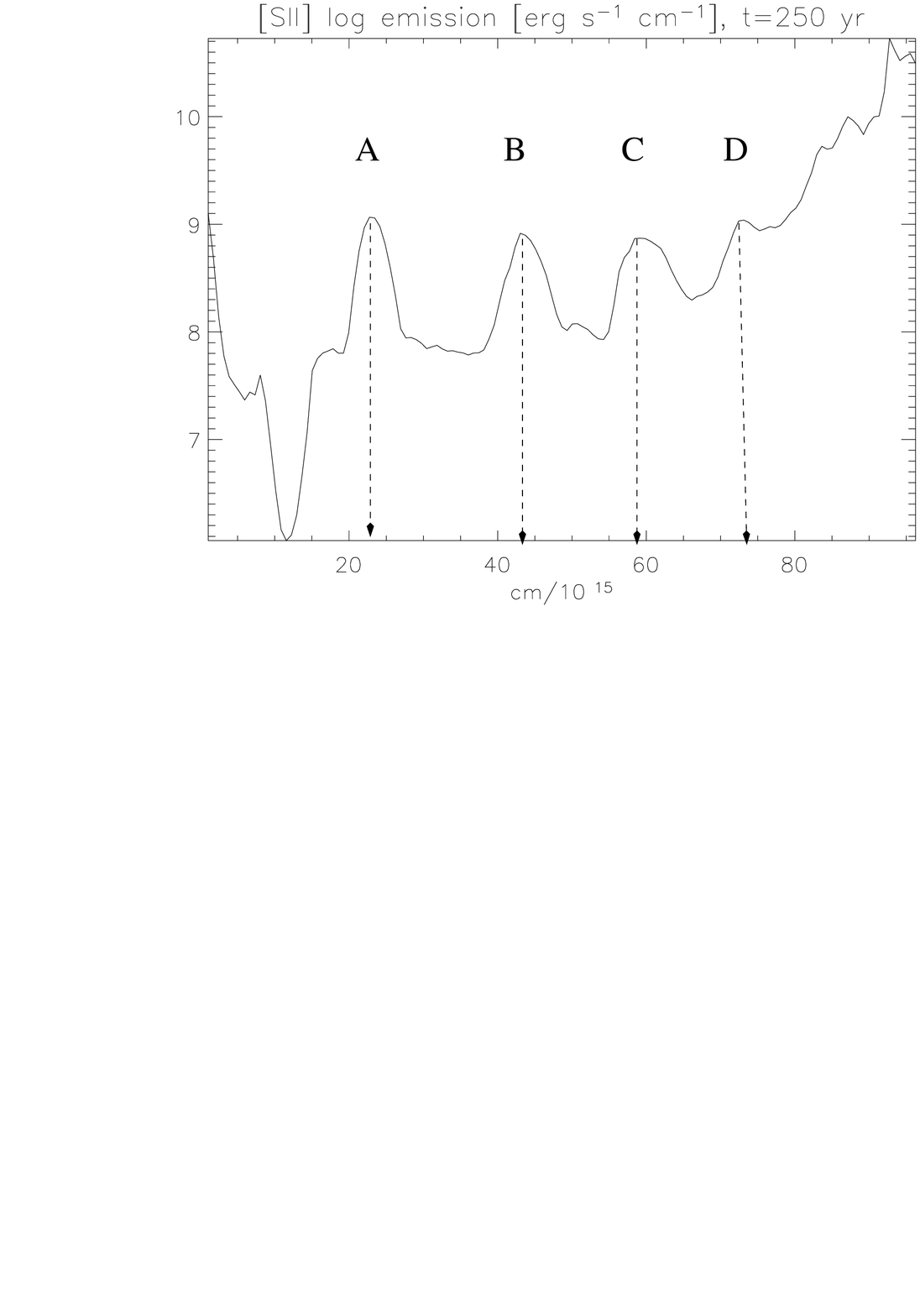} \\ 
\vskip -5cm
\includegraphics[height=11cm,width=8cm]{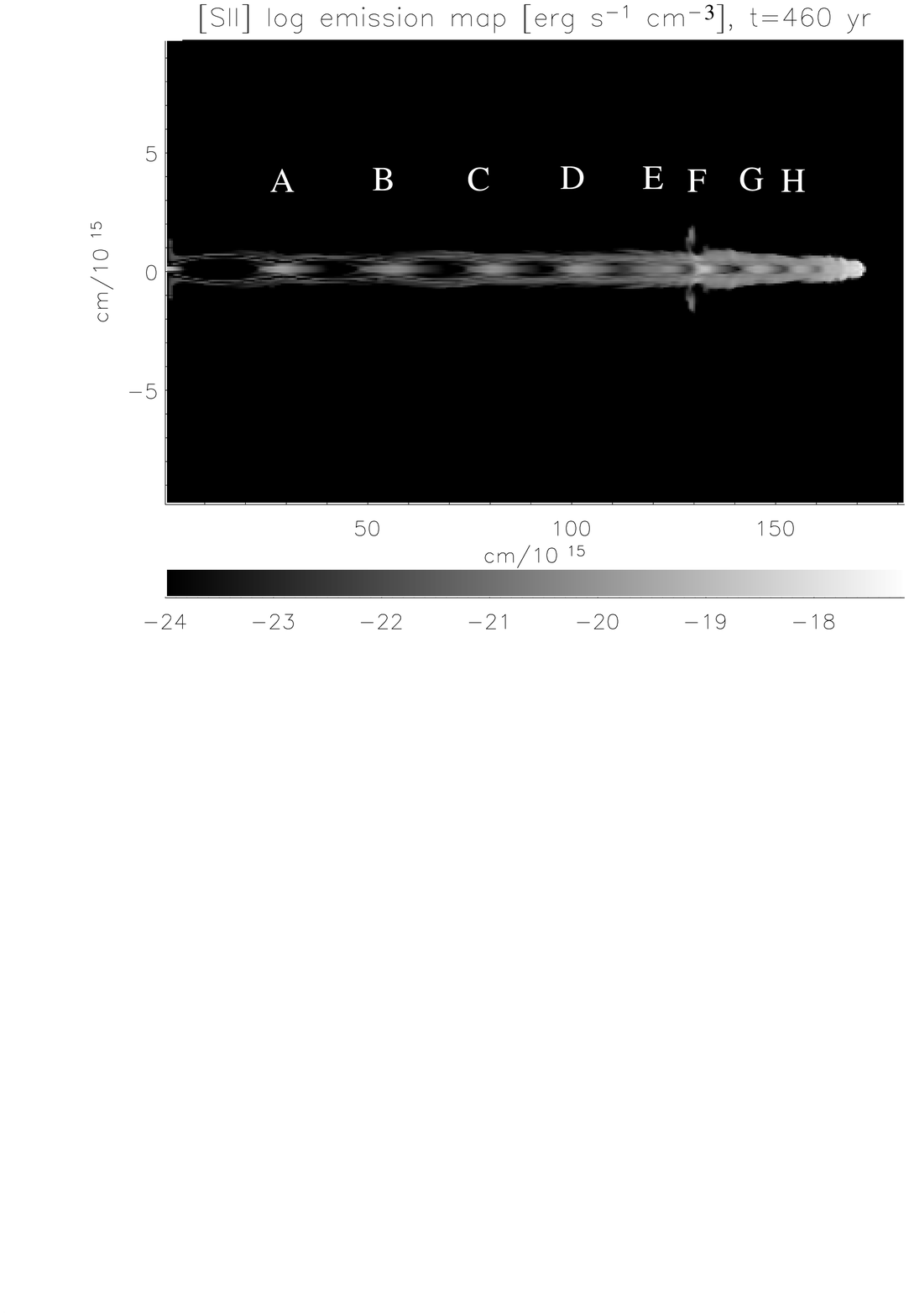}
\includegraphics[height=11cm,width=8cm]{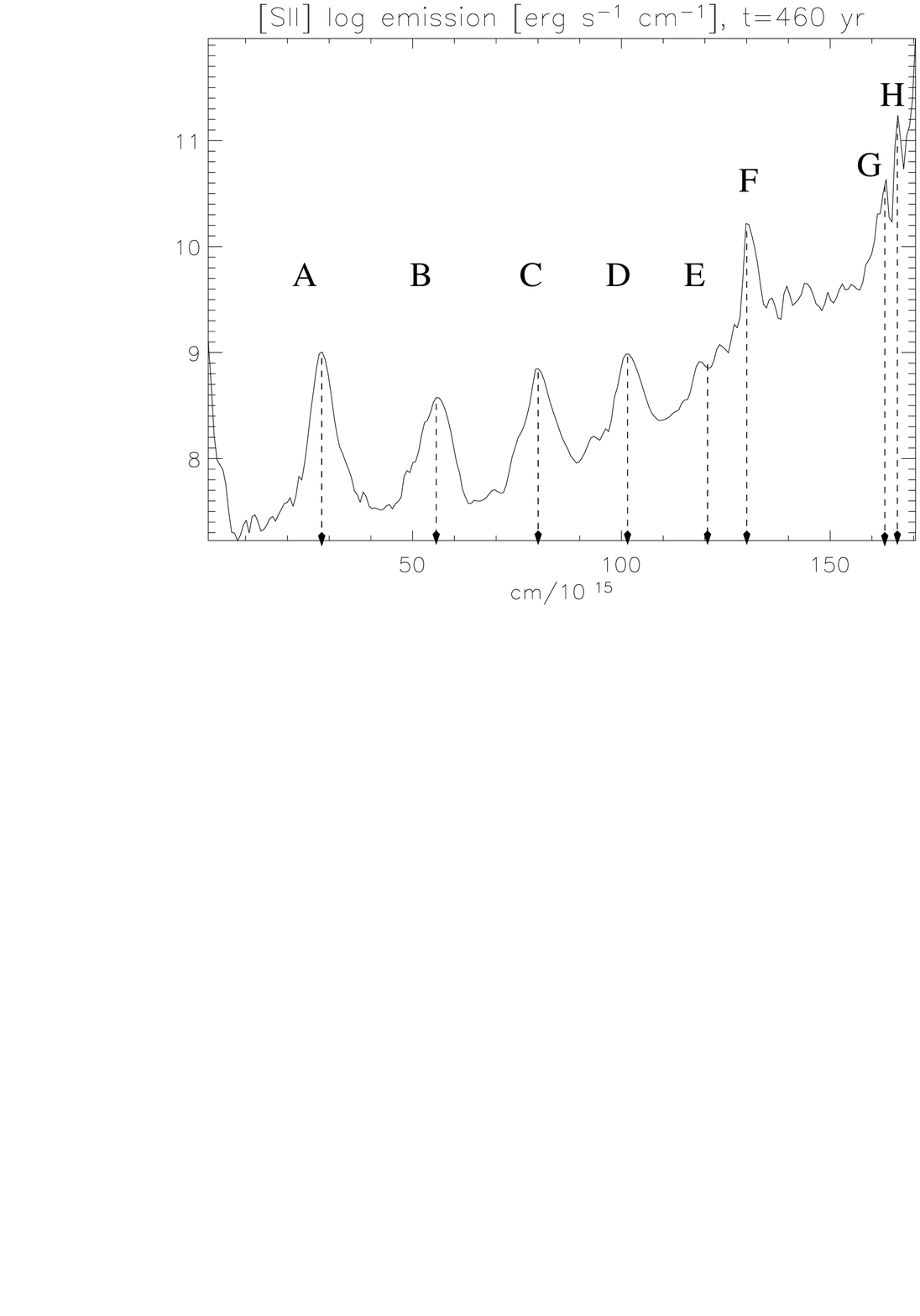} \\
\vskip -5cm
\includegraphics[height=11cm,width=8cm]{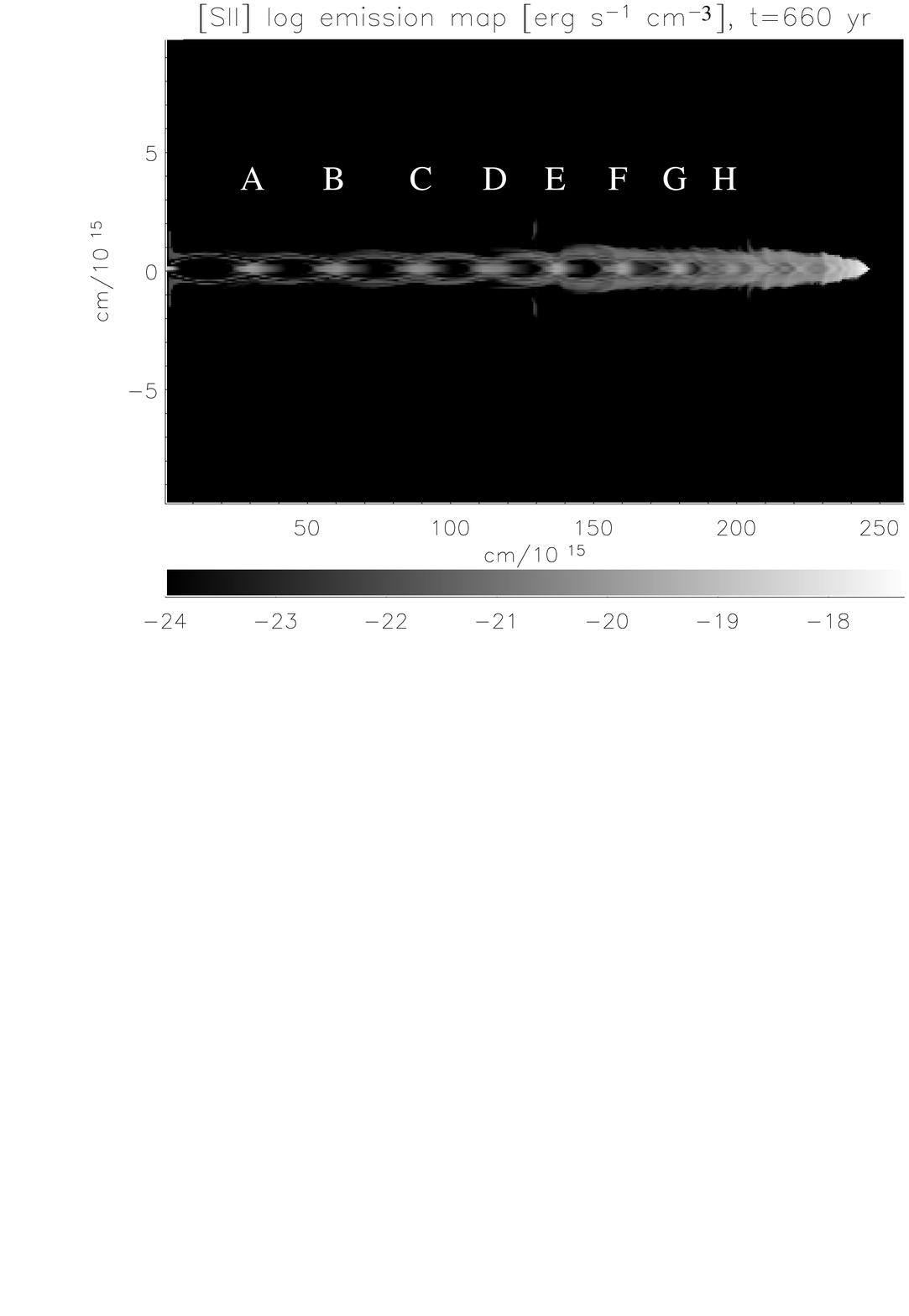}
\includegraphics[height=11cm,width=8cm]{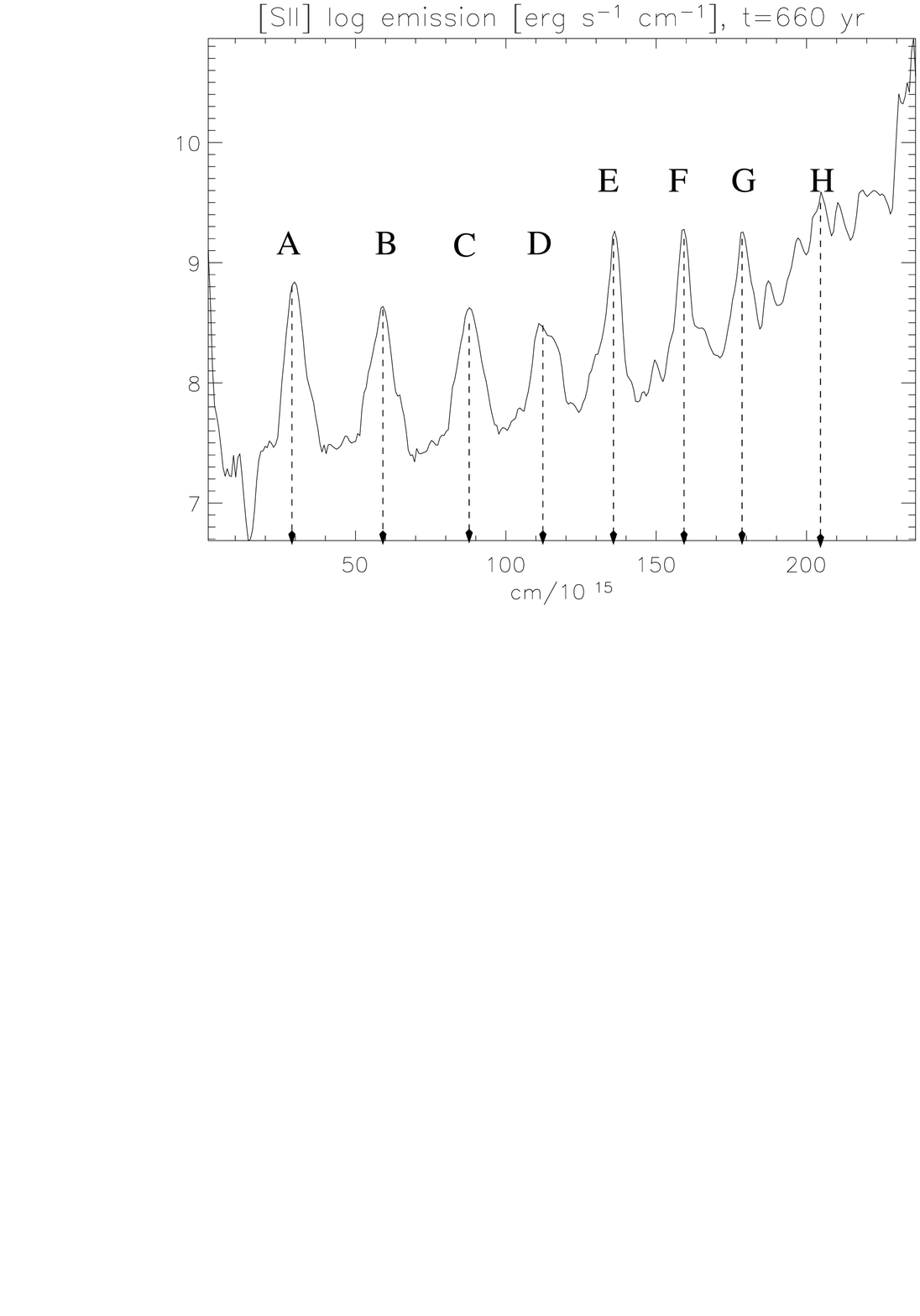} \\
\vskip -5cm
\caption{Case $\mathcal{D}$. [\ion{S}{II}] emission map (left panels) 
and emissivity function $E(\xi)$ (right panels) for different output
times.}
\label{fig:f9}
\end{figure*}

%
%
\begin{table}
\centering
\begin{tabular} {ccccccccccc}
\hline
\multicolumn{1}{c}{} &
\multicolumn{2}{c}{$t=135$~yr} &
\multicolumn{2}{c}{$t=250$~yr} &
\multicolumn{2}{c}{$t=460$~yr} &
\multicolumn{2}{c}{$t=660$~yr} \\
\cline{2-9}
knot & $\xi$ & $v$ & $\xi$ & $v$ & $\xi$ & $v$ & $\xi$ & $v$ \\ 
\hline
A & 16 & -- &  23 & 19 &  28 &  7 &   30 &  3\\
B & 30 & -- &  43 & 36 &  55 & 16 &   60 &  8\\
C & 41 & -- &  60 & 52 &  80 & 23 &   90 & 16\\
D & -- & -- &  73 & -- & 100 & 41 &  112 & 19\\
E & -- & -- &  -- & -- & 120 & -- &  137 & 27\\
F & -- & -- &  -- & -- & 130 & -- &  160 & 48\\
G & -- & -- &  -- & -- & 162 & -- &  180 & 29\\
H & -- & -- &  -- & -- & 165 & -- &  205 & 64\\
\hline
\end{tabular}
\caption{Case $\mathcal{D}$. Estimated position $\xi$ (in units 
of $\bar{L}=10^{15}$~cm) and velocity $v$ (in km~s$^{-1}$) along the axis
for the knots and output times of Fig.~\ref{fig:f9}.
}
\label{tab:2}
\end{table}

To allow for the formation of knots over the jet length scale 
and to reduce the intra-knot spacing, in case $\mathcal{D}$
the nozzle radius $r_\mathrm{jet}$ has been reduced to $0.1~\bar{L}$,
while $\eta$ has been increased to $0.4$,
preserving the usual inflow speed. The increase in density is needed
to keep a realistic mass injection rate, by (partially) compensating 
the loss of area. 
A spectacular chain of knots appears, very apparent in the  
[\ion{S}{II}] emissivity maps of Fig.~\ref{fig:f9} (left panels), at 
different output times of jet evolution. 
A secondary re-collimated jet appears at $\xi=130~\bar{L}$ (see the frame 
at $t=460$~yr), revealed by a light halo in the emissivity, 
corresponding to a high density ring of matter. 
As already mentioned in Sect.~\ref{sect:AB}, these \emph{re-collimation} 
effects are not real, but arise from numerical effects due to the 
symmetry of the geometry. Five knots are  
visible to the left of the re-collimation point (knots A, B, C, D, E), 
and three to the right (knots F, G, H). Velocities can be estimated by 
looking at the temporal evolution of the emission function $E(\xi)$, in the 
same figure (right panels).
The position of each knot can be identified by tracking the 
spatial variations of the corresponding peaks in the 1-D plots. 
Results are shown in Table~\ref{tab:2}.
The estimated velocities are seen to increase with the distance from 
the nozzle, from $3\mbox{ km s}^{-1}$ to $64\mbox{ km s}^{-1}$. Even
though these velocities are still far from what observed in real jets
(typically knots move at $\approx 70\%$ of the local flow speed), the trend
of increasing velocities with distance from the source is invariably
found in our simulations and it is a feature consistent with 
observations \citep{eisloeffel92}.

A further comparison with observations of real jets can be attempted
by measuring the decay in brightness of the knots over the beam length.
If we measure this quantity for the first knots for the final output
time $t=660$~yr, their brightness decays with an estimated power law 
exponent $\alpha\approx - 0.9$. This value does not match observations 
and theoretical models, which foresee $\alpha \approx -2$. 
A value of $\alpha=-1.9$ has been found, as an example, in HH~30 
\citep{ray96}.
In our simulation the value of $\alpha $ could be affected by the    
beam re-collimation, which changes the brightness slope downward
$\xi\simeq 130~\bar{L}$. This is why only the first four knots have
been considered in the present estimate. We believe that such a
problem will disappear in more realistic 3-D simulations.

\subsection{Restarting jets (case $\mathcal{E}$)}
\label{sect:restart}

%
%
\begin{figure*}
\centering
\includegraphics[height=11cm,width=8cm]{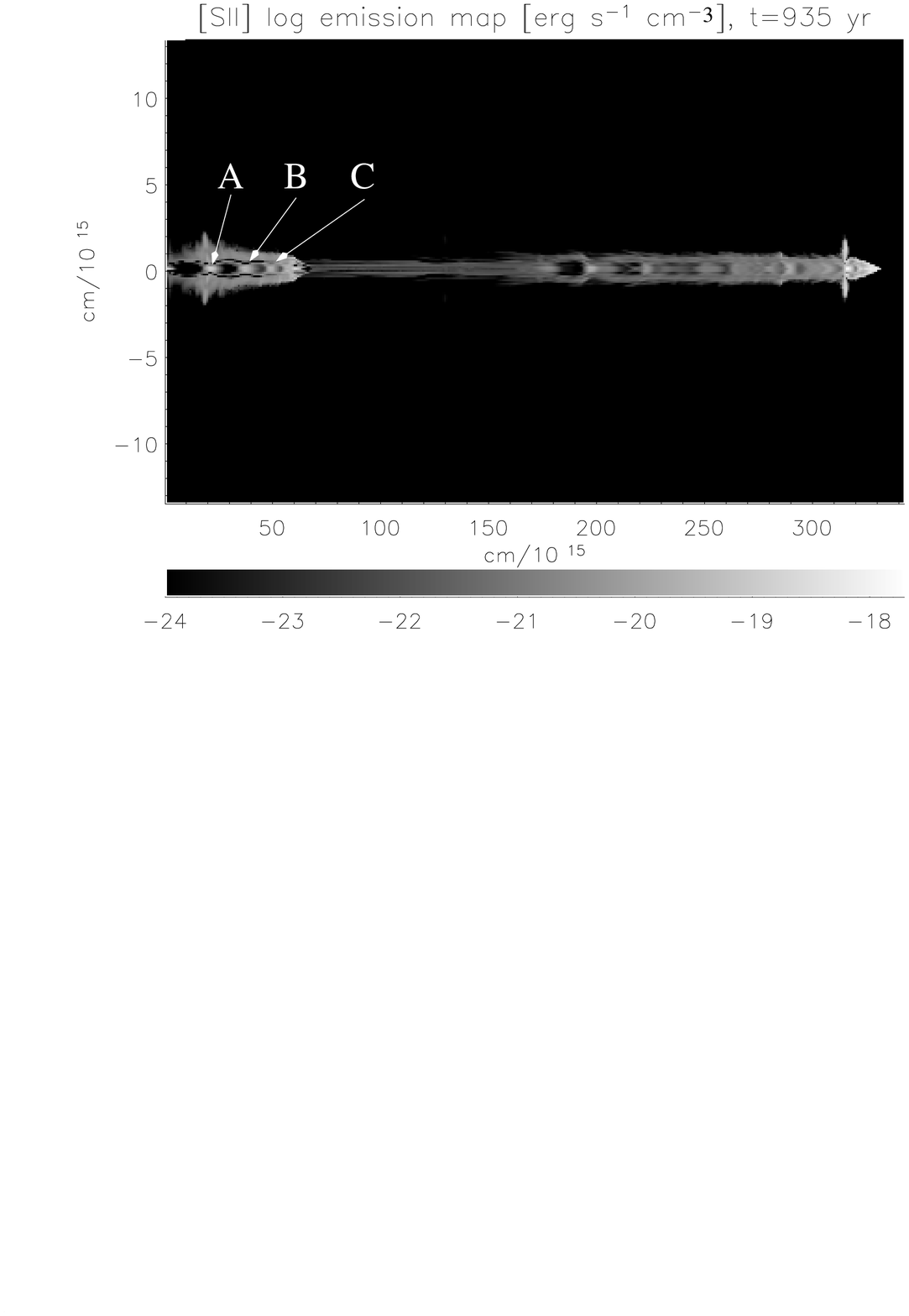}
\includegraphics[height=11cm,width=8cm]{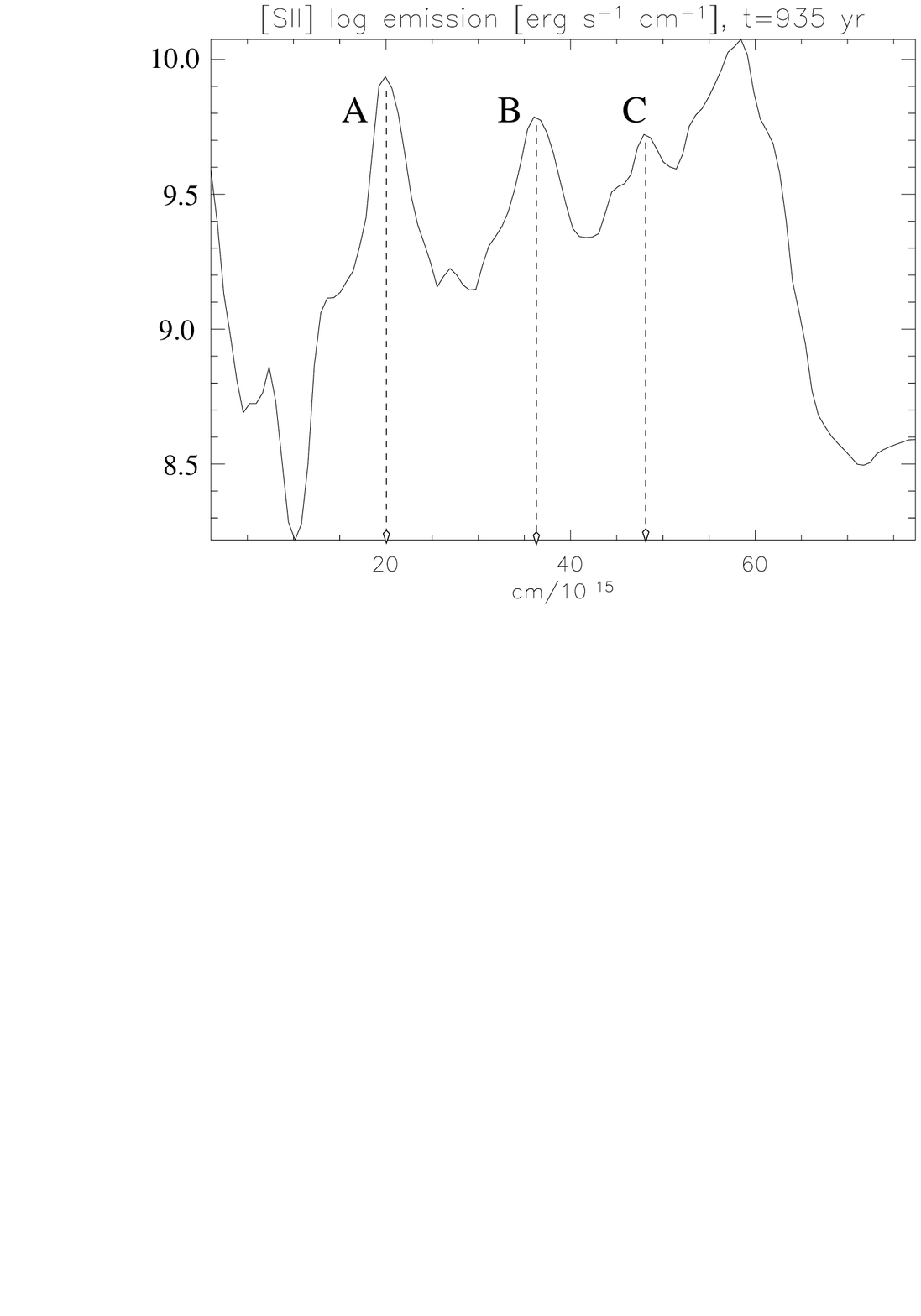} \\ 
\vskip -5cm
\includegraphics[height=11cm,width=8cm]{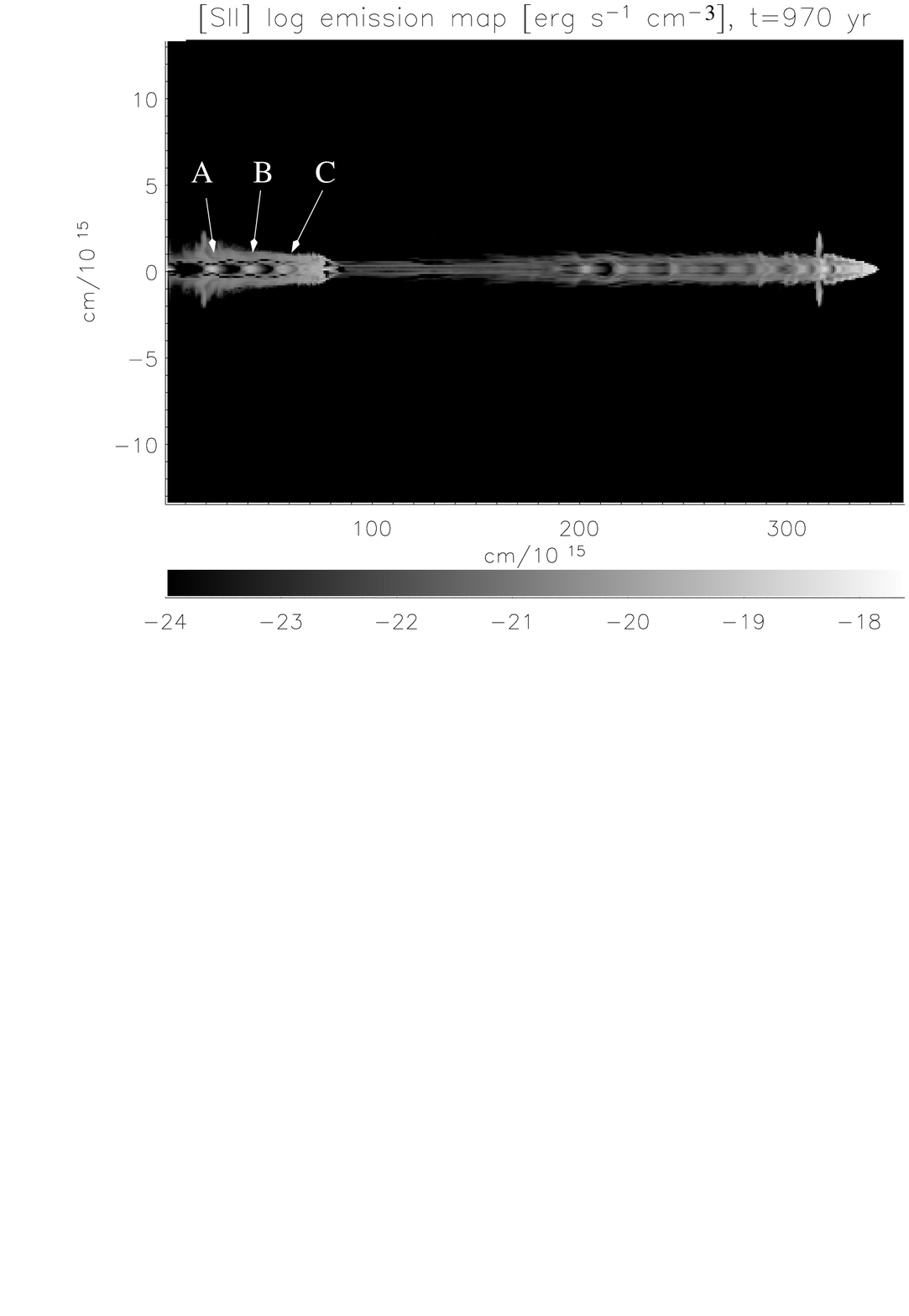}
\includegraphics[height=11cm,width=8cm]{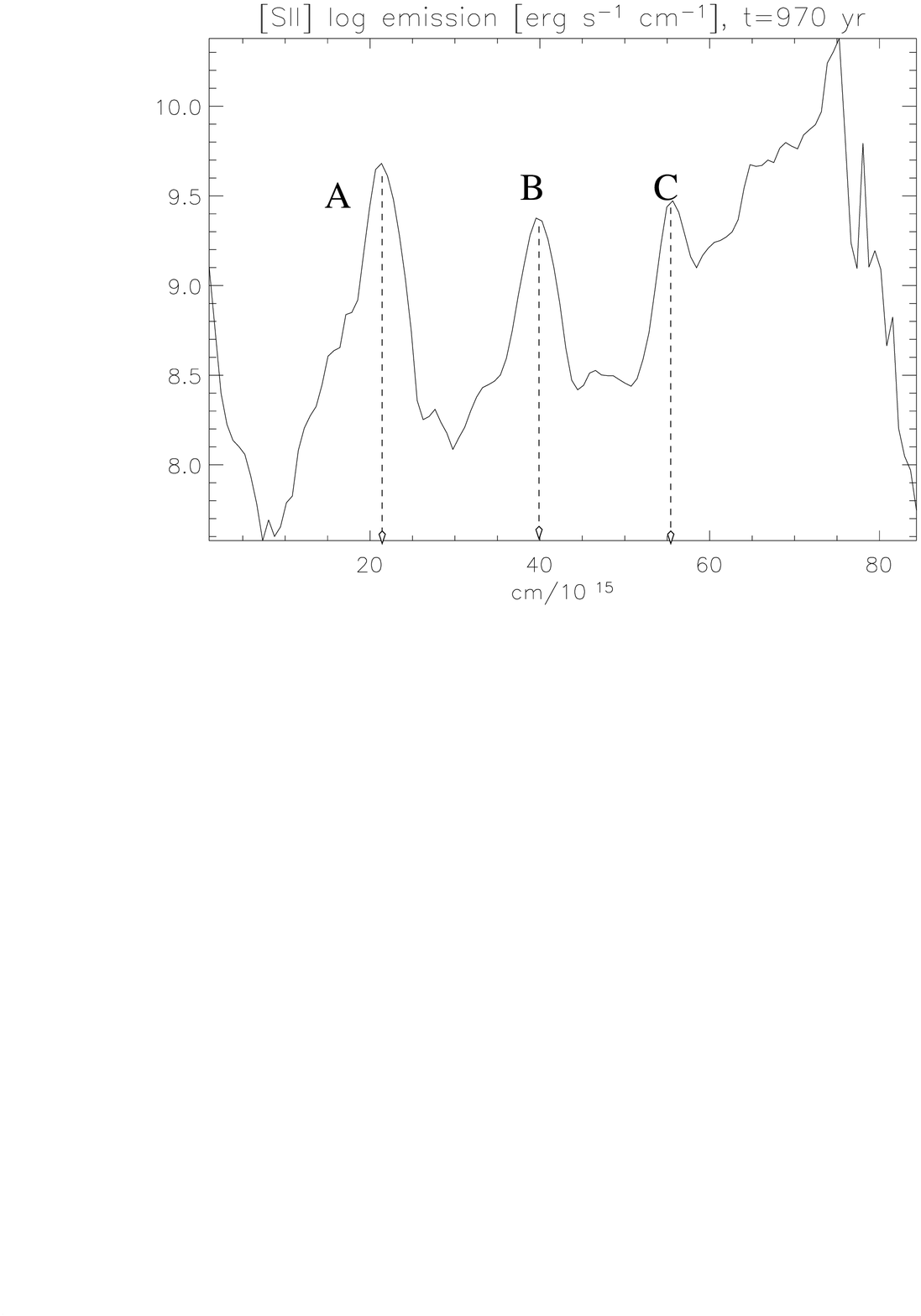} \\ 
\vskip -5cm
\includegraphics[height=11cm,width=8cm]{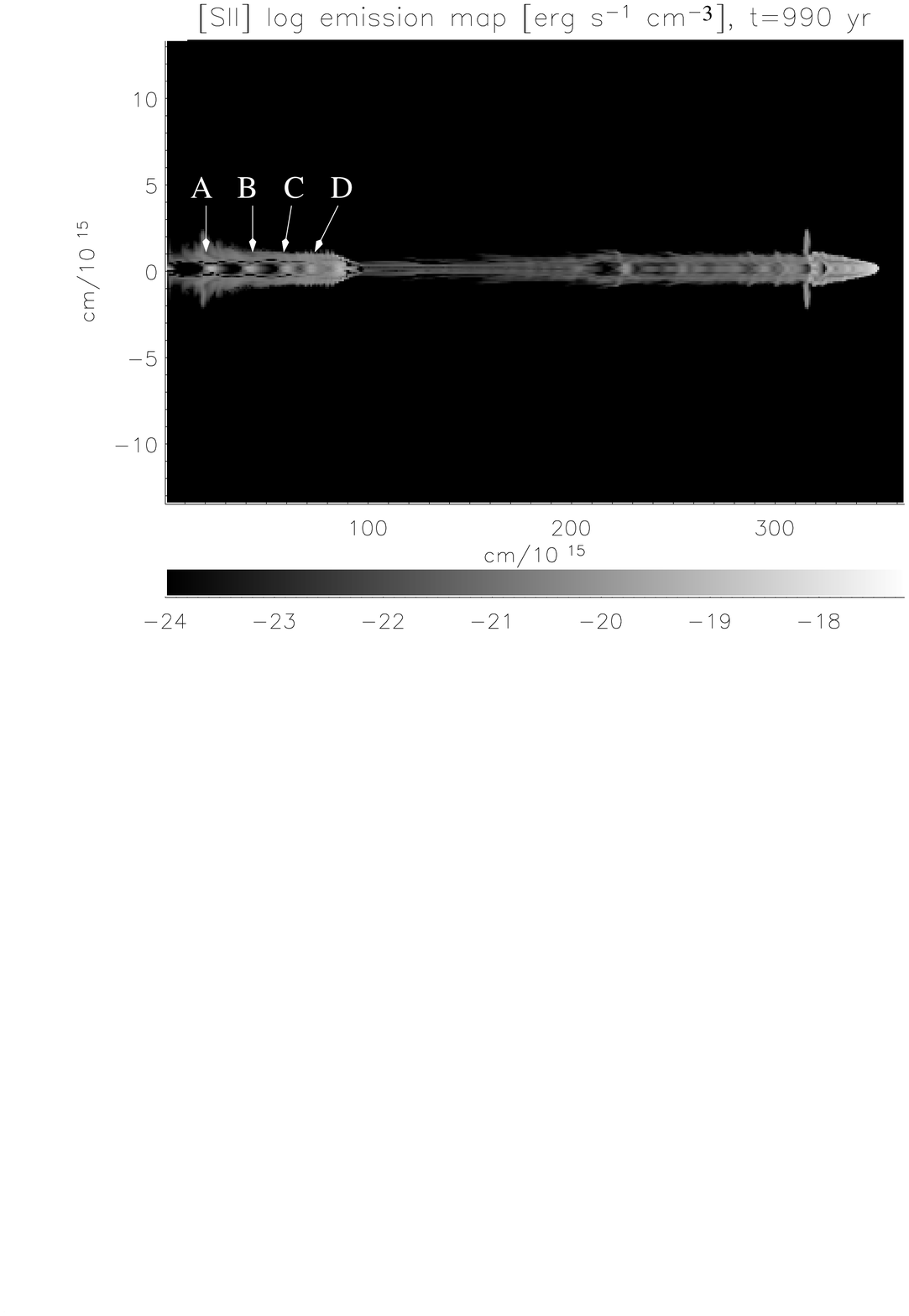}
\includegraphics[height=11cm,width=8cm]{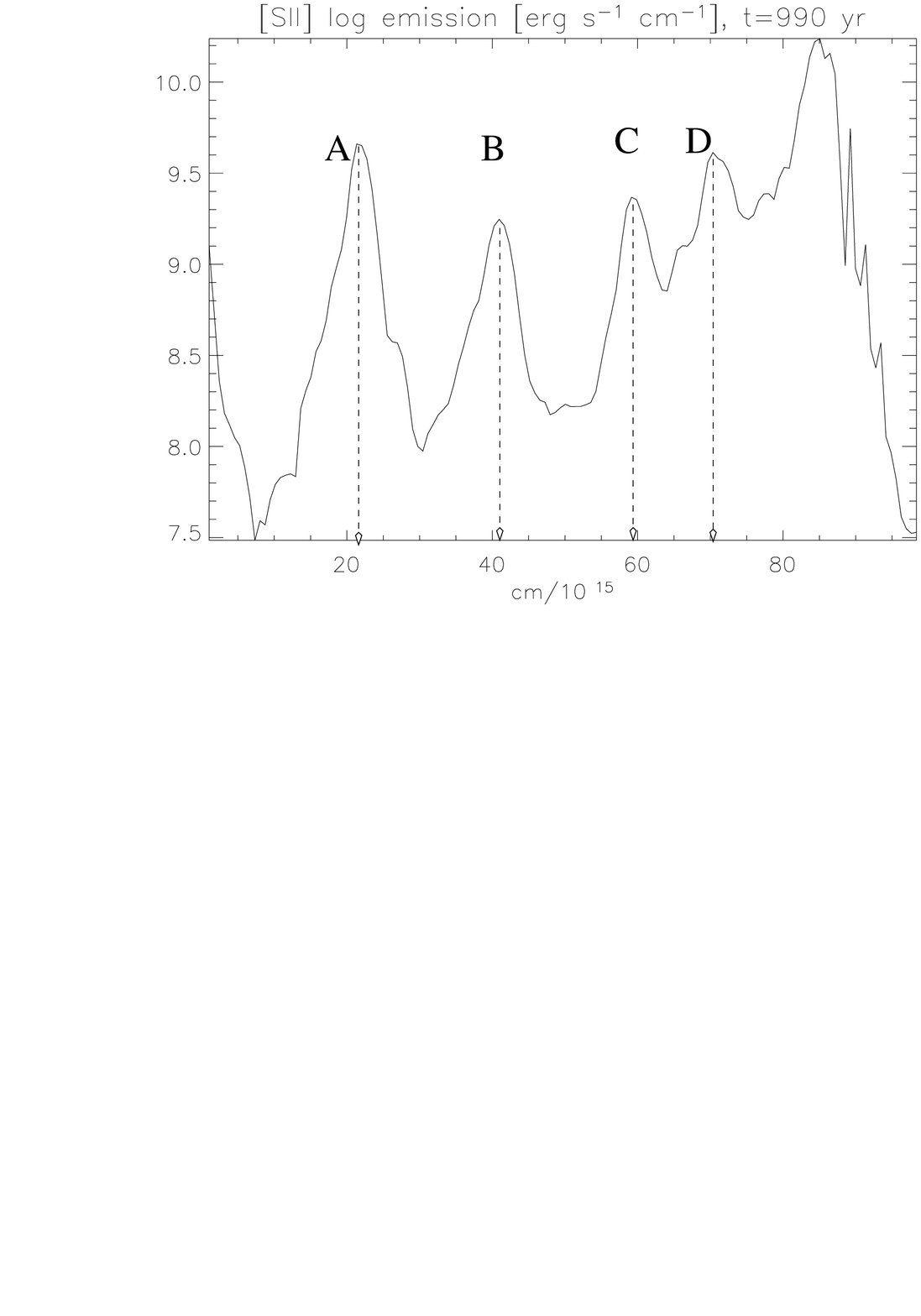} \\
\vskip -5cm
\includegraphics[height=11cm,width=8cm]{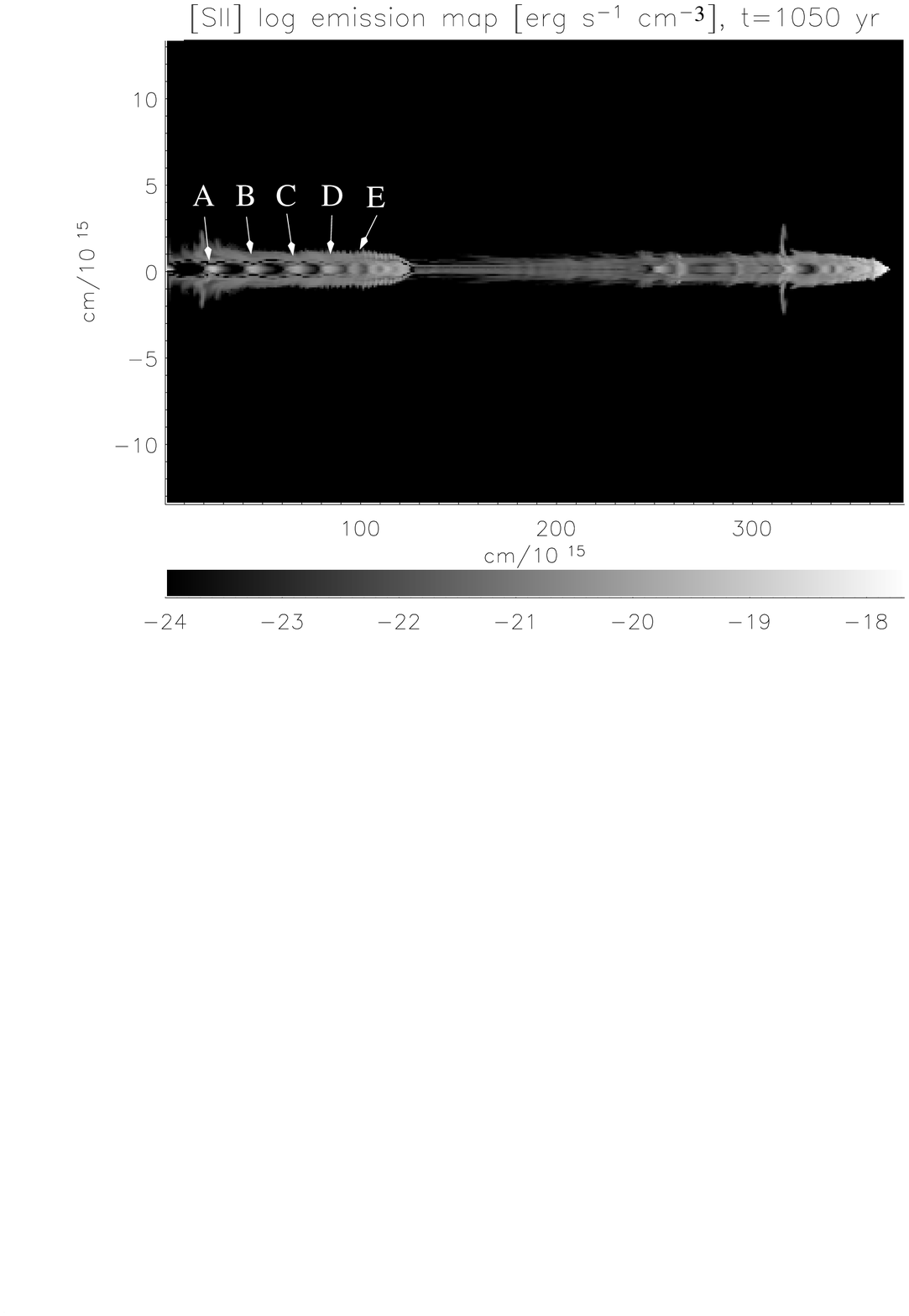}
\includegraphics[height=11cm,width=8cm]{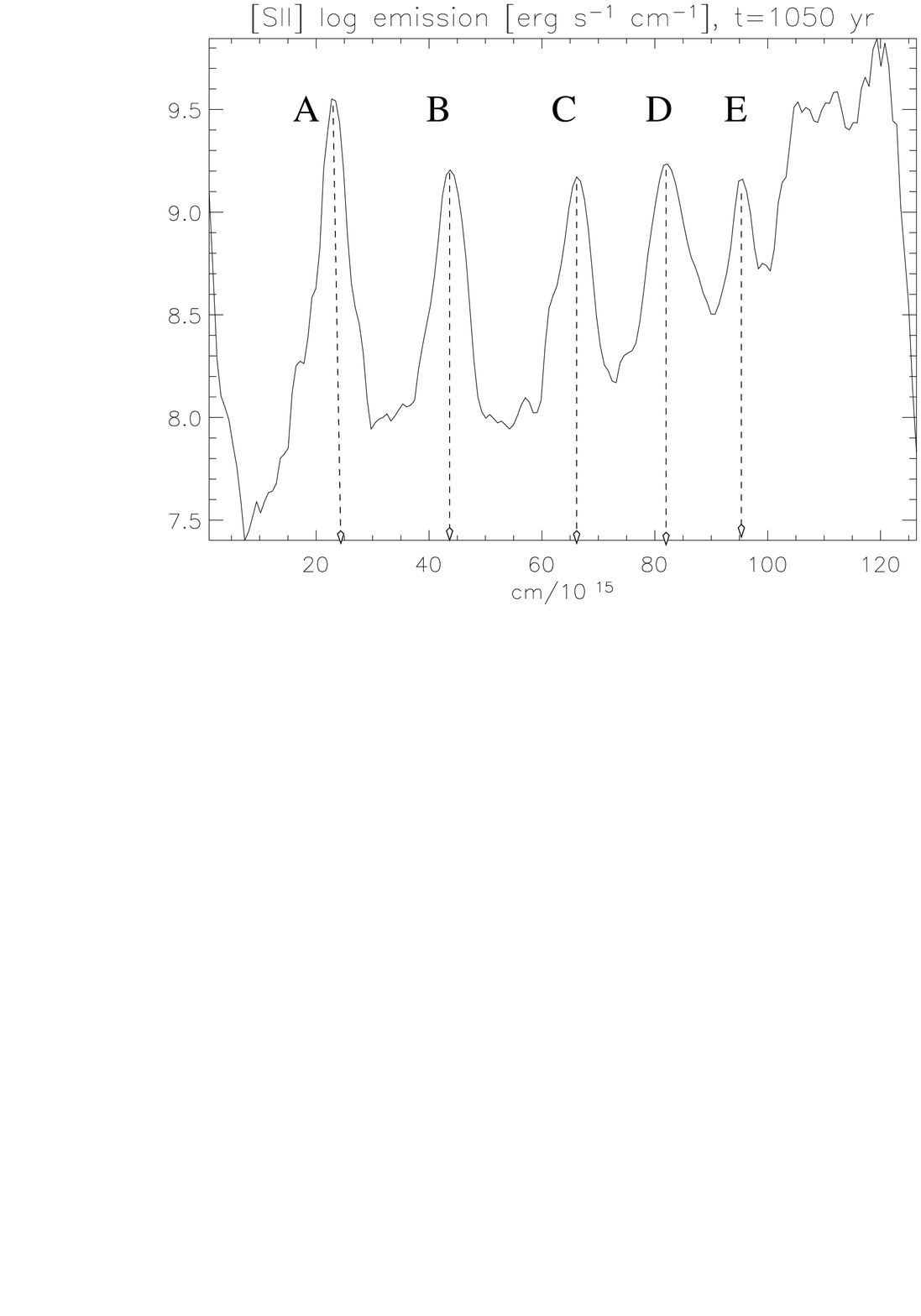} \\
\vskip -5cm
\caption{Case $\mathcal{E}$. [\ion{S}{II}] emission map and
emissivity function $E(\xi)$ at different output times.}
\label{fig:f10}
\end{figure*}

Case $\mathcal{E}$ retains the same parameters of the previous run. The 
difference is that a temporal discontinuity has been applied on the 
initial conditions at the nozzle.
Such time-varying initial conditions have nothing to do with the pulsating 
inflow conditions often invoked in the literature to generate moving knots.
In case $\mathcal{E}$ one makes the hypothesis that some  
discontinuity occurs at the jet source, namely that the flow at the nozzle 
is switched-off and that, after some time, it is switched-on again, so 
that former initial conditions are restored.
In this way a new, \emph{restarting jet} forms, which bores its way 
in the wake of the old one, producing new, interesting features.
The motivation for these inflow conditions is provided by the observed
formation of leading bow-shocks with temporal spacings of a few hundred years. 
In the present simulation a temporal variation has been set at the source 
of the jet with a time frequency small enough with respect to the intra-knot 
frequency, that one can still consider the overall structure as a 
steady jet. Note that this kind of variability is not meant 
to generate high-frequency internal working surfaces as in the pulsating 
inflow models, since in our simulations knots form because 
of IOS along the jet beam.  

The results of the simulation for case $\mathcal{E}$ are displayed
in Fig.~\ref{fig:f10}, where [\ion{S}{II}] 2-D maps and $E(\xi)$ plots
are shown at different output times of the jet evolution.
At $t=620$~yr the source of the jet has been thus switched off.
Both the inflow pressure and density, or mass 
injection rate, have been reduced by a factor of 10. 
At $t=720$~yr the original initial conditions 
have been restored, and a \emph{new} jet forms at the source and starts
to propagate in the wake of the old one.
In the displayed plots we report emission maps (left panels) and $E(\xi)$
functions (right panels), each row corresponding to different times
of the evolution of this two-jet system, from $t=935$ to $t=1050$~yr.
This choice of output times allows us to study in detail the new jet,
which exhibits a chain of knots quite close to the source,
a typical feature observed in many YSOs \citep[e.g.][]{reipurth01}.

%
%
\begin{table}
\centering
\begin{tabular} {ccccccccccc}
\hline
\multicolumn{1}{c}{} &
\multicolumn{2}{c}{$t=935$~yr} &
\multicolumn{2}{c}{$t=970$~yr} &
\multicolumn{2}{c}{$t=990$~yr} &
\multicolumn{2}{c}{$t=1050$~yr} \\
\cline{2-9}
knot & $\xi$ & $v$ & $\xi$ & $v$ & $\xi$ & $v$ & $\xi$ & $v$ \\ 
\hline
A & 20 & -- &  22 & 18 & 22 &  0 &  25 & 16 \\
B & 36 & -- &  40 & 36 & 41 & 16 &  44 & 16 \\
C & 48 & -- &  55 & 64 & 60 & 80 &  67 & 37 \\
D & -- & -- &  -- & -- & 70 & -- &  83 & 70 \\
E & -- & -- &  -- & -- & -- & -- &  95 & -- \\
\hline
\end{tabular}
\caption{Case $\mathcal{E}$. Estimated position $\xi$ (in units 
of $\bar{L}=10^{15}$~cm) and velocity $v$ (in km~s$^{-1}$) along the axis
for the knots and output times of Fig.~\ref{fig:f10}.
}
\label{tab:3}
\end{table}

Interesting results are found by estimating the velocity of such
knots (labeled A, B, C, D, E in Fig.~\ref{fig:f10}) 
from the positions of the corresponding peaks
in the integrated emission $E(\xi)$, as done in the previous sub-section. 
The kinematics of the new jet differs from the old one in two main aspects. 
The bow-shock propagates at a substantial fraction of the injection velocity 
$V_\mathrm{jet}=200\mbox{ km s}^{-1}$, while the external bow shock typically
propagates at a value which is about half of that speed. 
Moreover, the knots in the new jet are seen to move with
rather high individual velocities, as reported in Table ~\ref{tab:3}.
This different behavior with respect to $\mathcal{D}$ is due to 
the fact that the new jet travels in the low density wake of the old one, 
rather than in the higher density ambient of the unperturbed ISM. 
The most remarkable results of this particular experiment are then 
the morphology and kinematics of the new jet knots, which appear
close to the source and with relatively high-speed motions.

\section{Conclusions}

Our simulations show that under-expanded, light jets can \emph{naturally}
generate a pattern of emitting knots that possess proper motions, without 
invoking temporal variation of the source. In our scenario, knots are due 
to IOS, which are formed because of standard gas dynamical re-collimation
processes, and their proper motion is due to the interaction with a
highly time-dependent environment, namely the cocoon formed by the
propagation of the jet head. The resulting knots are seen to survive
radiative cooling and the synthetic images we obtained resemble
qualitatively the observations of many HH objects: the individual velocities
are seen to increase with distance from the source and the knots' brightness,
on the other hand, is found to decay over the beam length.
However, the detailed properties of such knots, in terms of brightness, 
position in space, proper motion, and intra-knot spacing, obviously
heavily depend on initial conditions. Exploring all, or even a large part, 
of parameter space  
is well beyond the scope and the possibility of this work, though 
some final considerations can be made:
\begin{itemize}
\item the fact that steady inflow conditions can drive the formation of 
propagating emitting knots is a remarkable result in itself, since 
it has often been argued that steady jets could only form steady 
internal structures; 
\item intra-knot spacing mostly depends on both $\Pi$, the pressure 
ratio and $r_\mathrm{jet}$, the nozzle radius. The correct choice of 
these parameters puts the knots length scale and the jet momentum loss
in a realistic range;
\item the velocities of the knots can reach a significant fraction 
of the jet bow-shock propagation speed, which, in turn, is typically 
half of the velocity along the beam (basically the injection velocity). 
In \emph{restarting} jets, simulating low-frequency (compared to the 
intra-knot frequency) variations of the inflow conditions, both the secondary
bow-shock and the newly born knots are seen to propagate faster, reaching
up to $40\%$ of the local flow speed, which is not too far from what is  
observed in real jets (about $70\%$). This promising branch of numerical 
experiments has just been opened and will be more exhaustively explored 
in the future;
\item the need for 3-D calculations arises from some significant discrepancies 
between numerical results and obervationally determined properties, 
such as the emissivity 
decaying exponent with the distance from the source. 
We claim here that such differences could 
arise from artificial re-collimation effects due to the assumed hypothesis
of axisymmetry, but 3-D simulations are needed to prove this statement.
\end{itemize}
  
In conclusion, our results show that IOS provide 
a natural, efficient mechanism for the formation of radiatively emitting 
knots which possess most of the observed features, such as proper motions
with increasing velocities along the jet beam.
However, claiming that they are the only driving mechanism is not realistic,
and most probably IOS work in co-operation with other mechanisms. 
In this framework, knots arising from IOS or from local working surfaces 
generated by inflow conditions fluctuations
could either co-exist or work separately in different objects.

\begin{acknowledgements}
This work was supported by the INAF (COFIN projects 2002 and 2004),
by the ASCI Flash Center, The University of Chicago,  
and by the CINECA Supercomputing Center. It was also
supported in part by the European Community's
Marie Curie Actions - Human Resource and Mobility within the JETSET (Jet 
Simulations, Experiments and Theory) network under contract MRTN-CT-2004 
005592.
The authors wish to thank Carlo Giovanardi for his help and
for fruitful discussions, Claudio Chiuderi for his useful 
and friendly suggestions, Simone Landi, Guido Bartoli 
(Scienza Industria e Tecnologia) for technical help, and, 
last but not least, an anonymous referee for his comments 
which helped to improve the manuscript.
\end{acknowledgements}

\bibliographystyle{aa}

\end{document}